\documentclass[sigconf]{acmart}

\usepackage{amsmath,amsfonts}
\usepackage{algorithmic}
\usepackage{textcomp}
\usepackage{xcolor}
\usepackage{booktabs}
\usepackage{array}
\usepackage{multirow}
\usepackage{multicol}
\usepackage{enumitem}
\usepackage{subfig}
\usepackage{bm}

\newcommand{\Mat}[1]{\bm{#1}}
\newcommand{\Vector}[1]{\bm{#1}}
\newcommand{\Set}[1]{\mathcal{#1}}
\newcommand{\Real}{\mathbb{R}}
\newcommand{\Loss}{\mathcal{L}}
\newcommand{\Normal}[1]{\mathrm{#1}}

\AtBeginDocument{%
  \providecommand\BibTeX{{%
    \normalfont B\kern-0.5em{\scshape i\kern-0.25em b}\kern-0.8em\TeX}}}

\copyrightyear{2021}
\acmYear{2021}
\setcopyright{acmcopyright}
\acmConference[CIKM '21]{Proceedings of the 30th ACM International Conference on Information and Knowledge Management}{November 1--5, 2021}{Virtual Event, QLD, Australia}
\acmBooktitle{Proceedings of the 30th ACM International Conference on Information and Knowledge Management (CIKM '21), November 1--5, 2021, Virtual Event, QLD, Australia}
\acmPrice{15.00}
\acmDOI{10.1145/3459637.3482443}
\acmISBN{978-1-4503-8446-9/21/11}

\begin{document}

\title{Learning Dual Dynamic Representations on Time-Sliced User-Item Interaction Graphs for Sequential Recommendation}

\author{Zeyuan Chen$^{1}$, Wei Zhang$^{1*}$, Junchi Yan$^{2}$, Gang Wang$^{3}$, Jianyong Wang$^{4}$}
\thanks{*Corresponding author.}
\affiliation{%
  \institution{$^{1}$East China Normal University, $^{2}$Shanghai Jiao Tong University, $^{3}$Beijing Institute of Technology, $^{4}$Tsinghua University}
  \country{}
}
\email{chenzyfm@outlook.com, zhangwei.thu2011@gmail.com}
\email{yanjunchi@sjtu.edu.cn, gangwang@bit.edu.cn, jianyong@tsinghua.edu.cn}

\renewcommand{\shortauthors}{Chen and Zhang, et al.}
\renewcommand{\shorttitle}{Dual Dynamic Representations for Sequential Recommendation}

\begin{abstract}
Sequential Recommendation aims to recommend items that a target user will interact with in the near future based on the historically interacted items.
While modeling temporal dynamics is crucial for sequential recommendation, most of the existing studies concentrate solely on the user side while overlooking the sequential patterns existing in the counterpart, i.e., the item side.
Although a few studies investigate the dynamics involved in the dual sides, the complex user-item interactions are not fully exploited from a global perspective to derive dynamic user and item representations.
In this paper, we devise a novel Dynamic Representation Learning model for Sequential Recommendation (DRL-SRe).
To better model the user-item interactions for characterizing the dynamics from both sides, the proposed model builds a global user-item interaction graph for each time slice and exploits time-sliced graph neural networks to learn user and item representations.
Moreover, to enable the model to capture fine-grained temporal information, we propose an auxiliary temporal prediction task over consecutive time slices based on temporal point process.
Comprehensive experiments on three public real-world datasets demonstrate DRL-SRe outperforms the state-of-the-art sequential recommendation models with a large margin.
\end{abstract}

\ccsdesc[500]{Information systems~Recommender systems}
\ccsdesc[300]{Information systems~Personalization}
\ccsdesc[100]{Information systems~Temporal data}

\keywords{sequential recommendation, user behavior analysis, graph neural networks, temporal point process}

\maketitle

\section{Introduction}\label{sec:intro}
Sequential recommendation~\cite{QuadranaCJ18} aims to recommend items that a target user prefers to interact with in the near future based on the interacted items in the past.
It has become a paradigmatic task in recommender systems in recent years.
This owes to the fast-developing online services (e.g., e-commerce platforms and streaming media) wherein user sequential behaviors are ubiquitous, and mature information technologies that make the collection of the sequential behaviors become easier. 
Compared to conventional recommendation settings~\cite{ShiLH14} that form user-item interaction records into a time-independent matrix and model each user-item pair separately, sequential recommendation is more accordant with real situations.
It could flexibly adjust recommendation results with the emergence of users' latest behaviors.

In the literature of sequential recommendation, there are roughly two branches of studies: session-based recommendation~\cite{hidasi2015session} and user identity-aware sequential recommendation~\cite{quadrana2017personalizing}.
For the former branch, it tends to consider users' short behavior sequences in recent history.
Usually, user identities are not available for session-based recommendation.
This is because in short sessions, users may not log onto accounts.
As a result, the majority of studies in this regard focus on building models with item sequences~\cite{KoukiFVCLJ20}.
For user identity-aware sequential recommendation, it is accustomed to taking long-term behavior sequences.
And user identity (ID) is commonly assumed to be known for model construction. 
This paper concentrates on the latter branch for leveraging users' long-term behaviors, which is more promising to pursue better recommendation performance.
Unless otherwise specified, we use sequential recommendation to represent user identity-aware sequential recommendation throughout this paper.

Sequential recommendation has been extensively studied in recent years~\cite{quadrana2017personalizing,FengLZSMGJ18,KangM18,PiBZZG19,LianWG0C20}.
Most of the studies share the same spirit that models the change of behavior sequences to characterize the dynamic user interests and learn the corresponding user representations.
As a supplement, user IDs are usually mapped to the same embedding space to enhance user representations.
Although the temporal dynamics in the user side are largely investigated, item representations are often assumed to be static by existing studies.
As such, the sequential patterns involved in the item side are often overlooked.
For example, down jackets are bought by more and more users with the coming of winter.
Grasping such popularity trends of items is hopeful for deriving better item representations, which would bring a positive effect on sequential recommendation.

To model the temporal dynamics in both the user and item sides, only a very few studies have been conducted~\cite{WuABSJ17,WuGGWC19,QinRF-WSDM20}.
They share a similar workflow that two deep sequential models (e.g., recurrent neural networks) are built for the user and item side, respectively.
The user behavior sequence (consisting of interacted items) and the item interaction sequence (consisting of interacted users) are commonly taken as model input to obtain their dynamic representations corresponding to different time steps/slices.
Although performance improvements could be achieved as compared to the sequential recommendation models only considering user dynamics, the workflow still encounters one major limitation: the abundant interactions between users and items happened in the past are often neglected and not explicitly modeled in a global perspective.
This will inevitably cause the user and item representations in user behavior sequences and item interaction sequences to be sub-optimal, which further affect the target user and item representations.

\begin{figure}[!t]
    \centering
	\includegraphics[width=1.\linewidth]{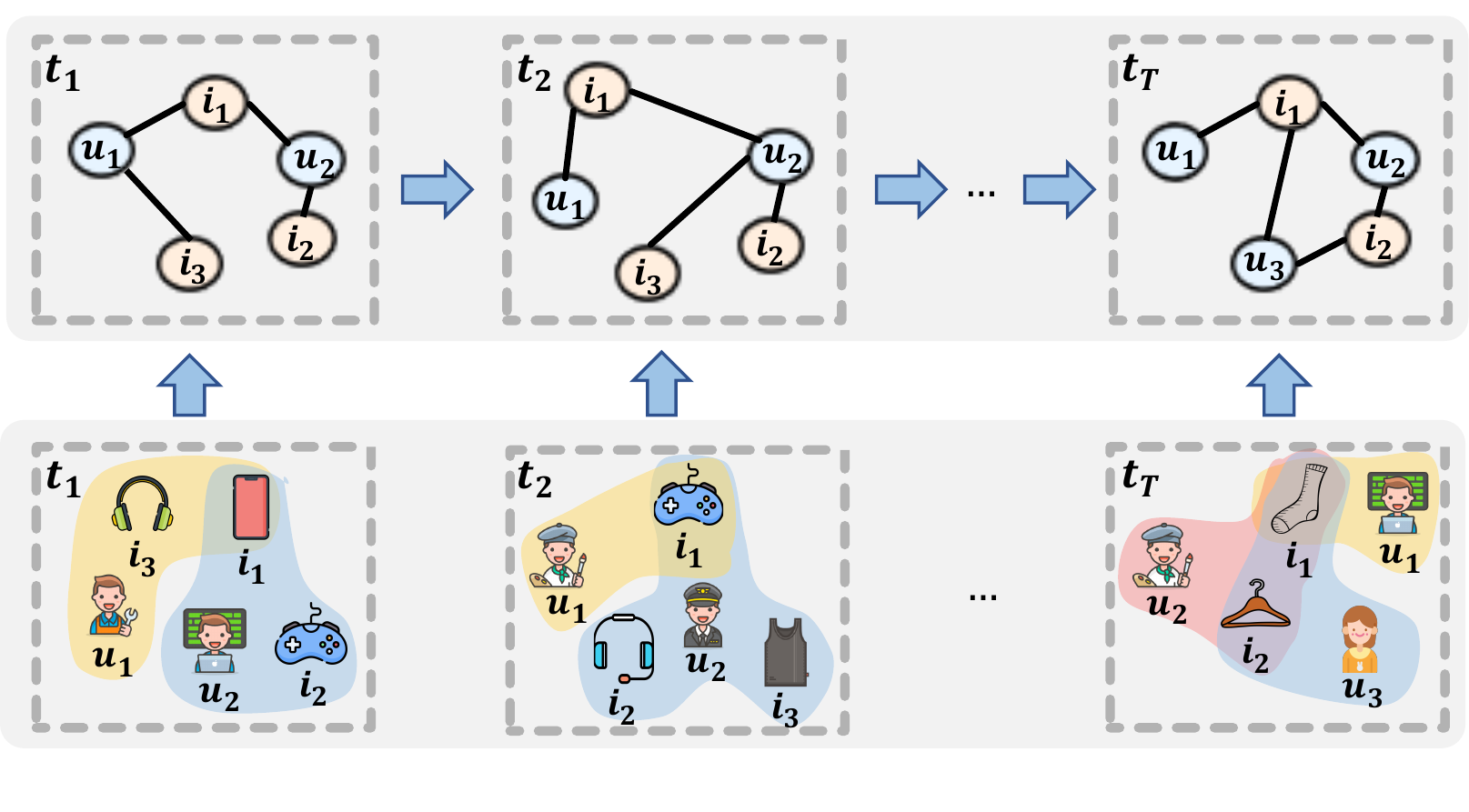}
	\vspace{-1em}
    \caption{Sketch for the time-sliced user-item interaction graph construction. Each dashed box corresponds to a time slice. It is worth noting that the numbers of users/items are not exactly the same for different time slices since not all users and items occur in each time slice.}
    \label{fig:sketch-graph}
\end{figure}

In this paper, we devise a novel Dynamic Representation Learning model for Sequential Recommendation (DRL-SRe) to tackle the above limitation.
It relies on the segmentation of the whole timeline into a specified number of equal-length time slices. 
For each time slice, a global user-item interaction graph is built to take all the users and items occurring in this slice as nodes.
Therefore it enables us to distill knowledge from all the user-item interactions to benefit the user and item representation learning.
The sketch for the time-sliced user-item interaction graph construction is shown in Figure~\ref{fig:sketch-graph}.

Given a sequence of the constructed user-item interaction graphs, DRL-SRe develops time-sliced graph neural networks over different time slices to propagate node representations so as to get user and item representations w.r.t. each time slice.
To correlate the time-sliced representations belonging to the same user or item, DRL-SRe utilizes two recurrent neural networks (RNNs) for the user side and the item side, respectively.
As such, the graph neural networks of different time slices are integrated into a unified model architecture.
To alleviate the issue of temporal information loss caused by the timeline segmentation, we further propose an auxiliary temporal prediction task over consecutive time slices based on temporal point process~\cite{Durbin1971Spectra}.
This empowers DRL-SRe with the ability to perceive the fine-grained temporal information and compensates for the primary sequential recommendation task.
In the end, DRL-SRe generates the possibility of the interaction between the target user and item in the near future.

We summarize the main contributions of this paper as follows:
\begin{itemize}[leftmargin=*]
    \item To our best knowledge, we are the first study to propose time-sliced graph neural networks for sequential recommendation, which is able to model the abundant and high-order user-item interactions from a whole perspective to obtain better dynamic user and item representations.
    
    \item In order to enable DRL-SRe to capture fine-grained temporal information, we integrate temporal prediction over consecutive time slices into the training of the proposed model and show that the auxiliary task can obviously promote the sequential recommendation task. 
    
    \item Extensive evaluations on three public real-world datasets have demonstrated the proposed model significantly outperforms the state-of-the-art sequential recommendation models and validated the contributions of its crucial components.
    
\end{itemize}

\section{Related Work}\label{sec:related}
This section concretely discusses the sequential recommendation studies and the relevant techniques for recommendation, i.e., graph neural networks and temporal point process.

\subsection{Sequential Recommendation}
Compared to session-based recommendation~\cite{KoukiFVCLJ20}, sequential recommendation studied in this paper considers user IDs and their behavior sequences in a longer time period.
The early work~\cite{ZhangDXFWBWL14} proposes to learn users' whole feature sequences by recurrent neural networks, wherein user IDs are simply taken as input.
On the contrary, Quadrana et al.~\cite{quadrana2017personalizing} segmented the whole user behavior sequences into multiple short sessions and adopted a hierarchical recurrent neural network to integrate these sessions.
To discriminate the short-term user dynamics and long-term user preference, the studies~\cite{FengLZSMGJ18,LiZLHMC18} first learn long- and short-term representations and then fuse them for prediction. 
Beyond the above schemes for handling sequences, some recent studies investigate advanced mechanisms to obtain behavior sequence representations~\cite{KangM18}, such as self-attention mechanisms~\cite{VaswaniSPUJGKP17}, or apply sequential recommendation to specific scenarios where additional information (e.g., spatial information) is utilized~\cite{LianWG0C20}.
However, these studies do not consider the temporal dynamics in the item side.

Among the studies for sequential recommendation,~\cite{WuABSJ17,WuGGWC19,QinRF-WSDM20} are most relevant to this study.
The pioneering model RRN~\cite{WuABSJ17} utilizes two RNNs to learn the temporal dynamics of the user side and the item side, respectively.
The output representations of the last time step for the two RNNs are coupled to denote user preference and item features for computing the interaction probability.
DEEMS~\cite{WuGGWC19} shares a similar spirit of applying two RNNs to get user and item representations, and provides a new perspective of loss function construction. 
It not only adopts a user-centered sequential recommendation loss, but also introduces item-centered information dissemination loss to form an enriched loss function.
Since the above two methods loosely couple the two sequences and model them independently, SCoRe~\cite{QinRF-WSDM20} is proposed to leverage the past user-item interactions to facilitate the representation learning.
Nevertheless, the considered user-item interactions by SCoRe are limited to the direct neighbor nodes of a target user-item pair, thus lacking an effective manner to globally utilize all user-item interactions in a time slice.

It is worth noting that there are several studies~\cite{KumarZL19,ChangLW0FS020,LiZWLWY20} that learn user and item representations in a continuous time fashion.
Nevertheless, they are confined to the setting where the next user-item interaction time is required to be known in advance to derive real-time user and item representations, which is entirely different from this study.
Moreover, each user-item interaction in the history could only be modeled once under this situation, thus lacking an effective mechanism to capture high-order relations between users and items through repeated propagation over user-item interactions.

\subsection{Relevant Techniques for Recommendation}

\noindent \textbf{Graph Neural Networks.}
Graph Neural Networks (GNNs)~\cite{ZhouGNN18} have been undergoing rapid development currently and observing many applications in recommender systems.
Because user-item interactions are one kind of edges, some studies~\cite{Wang0WFC19,HeGCN2020} convert the commonly used user-item rating matrix into a user-item interaction graph and develop GNNs to act upon it.
Despite user-item interactions, some other studies design GNNs for modeling user social relations~\cite{Fan0LHZTY19,TangLSSMW20} and item relations in knowledge graphs~\cite{WangZWZLXG19}.
Besides, the scalability issue of GNNs in recommender systems is also treated.
For example, node sampling is used in representation propagation in graphs~\cite{YingHCEHL18}. 

In light of session-based recommendation where only user dynamics are considered, SR-GNN~\cite{sr_gnn} is the first GNN-based model that builds an item-item interaction graph for each session.
Inspired by this, three recent studies~\cite{WangZLLZLZ20,QiuHLY20,Wang0CLMQ20} have put forward to build a global item-item interaction graph over all sessions instead of each session.
Moreover, LESSR~\cite{ChenW20} proposes to preserve the order of interactions when using GNNs for each session.
However, very few studies have investigated GNNs for the studied sequential recommendation problem where both user and item dynamics should be modeled based on user-item interactions in time slices.
Although GLS-GRL~\cite{Wang0RQ0LZ20} models user-item interactions in the sequential scenario, the GNN-based model design serves for learning user group representations and the recommended items are towards user groups but not individuals.
In this paper, we develop time-sliced graph neural networks for modeling user-item interactions in multiple graphs to realize sequential recommendation.
Note that although a few studies~\cite{SankarWGZY20,LuoZYBYLQY20,WuCCH20} research on dynamic graph neural networks over graph snapshots/time slices, none of them have been applied to the scenario of sequential recommendation.

\noindent \textbf{Temporal Point Process.}
Temporal point process~\cite{Durbin1971Spectra} is an elegant mathematical tool for modeling asynchronous time series.
Its highlight is to capture the continuous temporal gaps compared to time discretization methods.
The last several years have witnessed its application to user behavior trajectories~\cite{DuDTUGS16,VassoyRSA19,ChenLCL20,Liang20TNNLS}.
Most of these studies adopt sequential models to achieve interaction prediction and time prediction simultaneously.
By contrast, this paper differs from them by conducting temporal prediction over the past consecutive time slices, hoping to mitigate the temporal information loss incurred by the time slices.

\section{Problem Formulation}\label{sec:problem}
In the domain of recommender systems, a matrix $\Mat{R} = \{r_{ui}\}_{M\times N}$ is usually defined to record user-item interactions.
We assume $u\in\Set{U}$ and $i\in\Set{I}$, and $M$ and $N$ represent the size of user set $\Set{U}$ and item set $\Set{I}$, respectively.
The values in the matrix are determined based on whether user feedback is explicit (e.g., rating) or implicit (e.g., click).
Here we focus on implicit feedback since it is more common and easily obtained than explicit feedback.
As such, we define $r_{ui} = 1$ if user $u$ has interacted with item $i$, and $r_{ui} = 0$ otherwise.
To adapt to sequential recommendation, each user-item interaction is associated with a timestamp $t$ to indicate when the interaction happens.
Accordingly, we have a triplet $(u,i,t)$ to denote one interaction and define $\Set{T} = \{(u,i,t)\}$ to cover all the observed interactions.

To facilitate the learning of dual dynamic representations based on graph neural networks, we segment the whole timeline into $T$ time slices with the equal-length time interval $\Delta T$.
Therefore we have $\Set{T}=\{\Set{T}^1,\Set{T}^2,...,\Set{T}^T\}$ where $\mathcal{T}^s$ contains all the observed triplets that occur in the $s$-th time slice.
It is worth noting that some users and items might not have any interaction in a time slice.
Under this situation, the corresponding user and item representations would keep unchanged in that time slice.
Now we proceed to formulate the sequential recommendation problem as follows:
\newtheorem{problem}{Problem}
\begin{problem}[Sequential Recommendation Problem] For a target user $u$, a candidate item $i$, and all the observed user-item interactions $\Set{T}$, the aim of the sequential recommendation problem is to learn a function $f$ that predicts their potential interaction probability in the near future, which is defined as
\begin{equation*}
    \hat{y}_{ui}=f(u, i, \Set{T};\Theta)\,.
\end{equation*}
\end{problem}

\section{Methodology}
\begin{figure*}[!t]
    \centering
	\includegraphics[width=.88\linewidth]{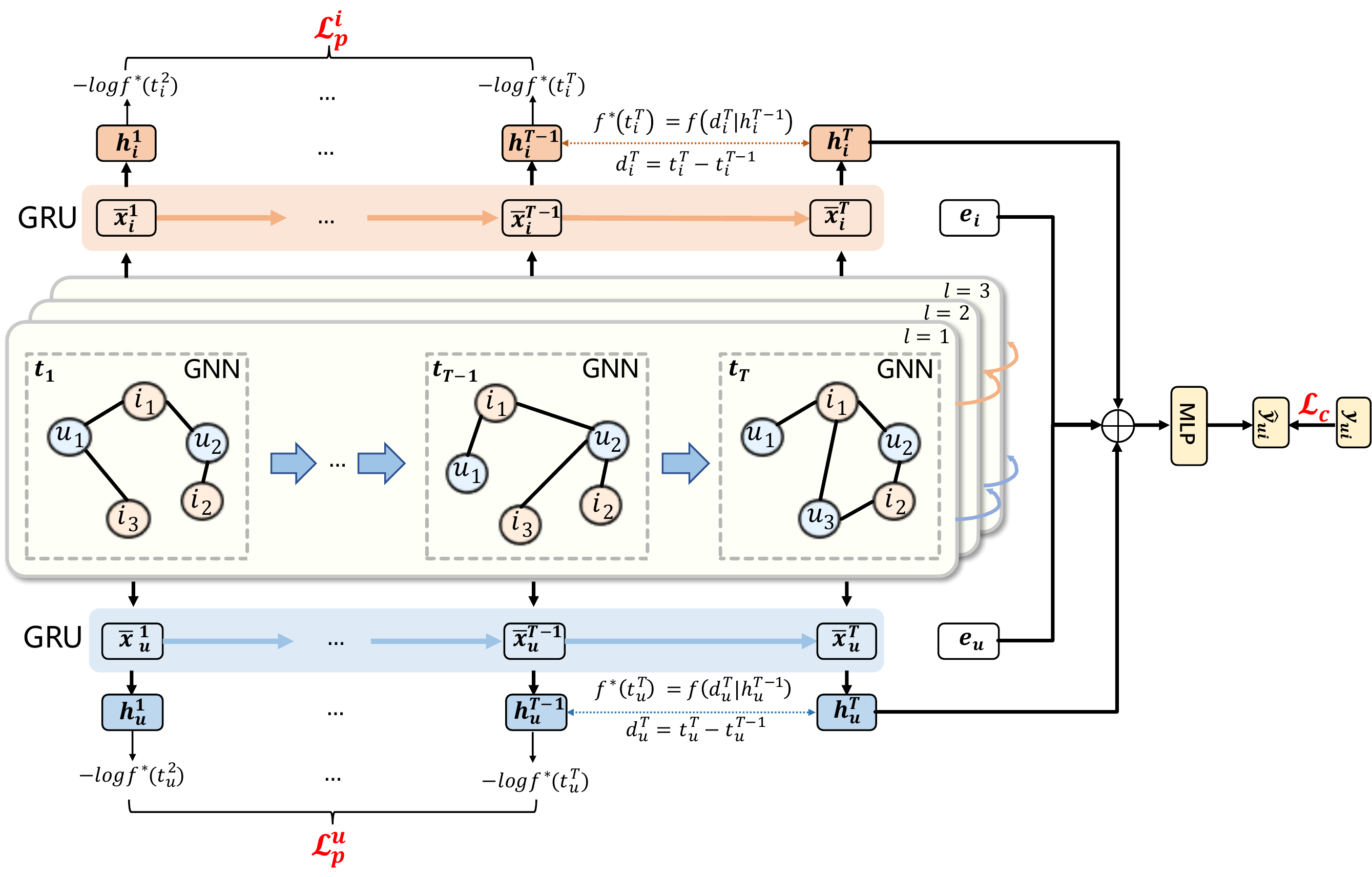}
	\vspace{-1em}
    \caption{Architecture of the model DRL-SRe, where the loss functions marked with red color are added.}
    \label{fig:model-architecture}
\end{figure*}

\textbf{Overview:} Figure~\ref{fig:model-architecture} depicts the overall architecture of the proposed model DRL-SRe, which takes all the observed user-item interactions as input and outputs the predicted interaction probability for a target user-item pair.
Within the model, time-sliced graph neural networks are proposed to learn two sequences of representations for the user and item sides.
Each representation in the two sequences corresponds to one time slice, facilitating to characterize the temporal dynamics.
Then the user and item representations of the current time slice, together with the static representations gotten from user and item IDs, are fed into the prediction part.
For the optimization part, the primary sequential recommendation loss $\Loss_c$ and the auxiliary temporal prediction loss $\Loss_p$, consisting of $\Loss_p^i$ for item $i$ and $\Loss_p^u$ for user $u$, are leveraged to jointly optimize the model end-to-end. 
In what follows, we organize the model specification in a logical fashion.

\subsection{Input Representations}
Following the convention of modern deep recommendation methods~\cite{ZhangYST19}, we map a user ID or an item ID to a dense vectorized representation.
This time-independent representation is thus static and used to reveal user long-term interests and item long-term characteristics.
Formally, taking user $u$ and item $i$ as examples, we define the following mapping functions:
\begin{equation}\label{eq:map}
\Vector{e}_u = \Mat{E}_U\Vector{o}_u\,,\quad\quad \Vector{e}_i = \Mat{E}_I\Vector{o}_i\,,
\end{equation}
where $\Mat{E}_U\in\Real^{d\times M}$ and $\Mat{E}_I\in\Real^{d\times N}$ are trainable embedding matrices for users and items, respectively.
$d$ denotes the embedding dimension.
$\Vector{o}_u$ is the one-hot encoding for user $u$ and it is analogous to $\Vector{o}_i$ of item $i$.

\subsection{Time-sliced Graph Neural Networks}
Before delving into the computational formulas of this module, we first clarify how to build user-item bipartite graphs $\Set{G}=\{\Set{G}^1,\Set{G}^2,...,\\ \Set{G}^T\}$ for each time slice.
Taking graph $\Set{G}^s$ for example, we build it based on all the user-item interaction records in $\Set{T}^s$.
Assume there are $M^s$ users and $N^s$ items occurring in the $s$-th time slice, then we have a node feature matrix $\Mat{X}^s\in\Real^{(M^s+N^s)\times d}$ and an adjacency matrix $\Mat{A}^s\in\{0,1\}^{(M^s+N^s)\times(M^s+N^s)}$ for the graph.
An entry in the adjacency matrix takes a value of 1 if the corresponding user-item interaction occurs, and 0 otherwise.
We do not take fine-grained edge weights since the ratios of repeated user-item interactions in a time slice are very small or even 0 for the used datasets.
However, we should emphasize that our graph could be easily generalized to take fine-grained edge weights if necessary.

The module of time-sliced graph neural networks is responsible for learning node representations based on the constructed graphs.
By convention, it performs representation propagation along with the edges of graph $\Set{G}^s$, which is defined as follows:
\begin{equation}\label{eq:propagation}
    \Mat{X}^s_{l+1} = \Mat{\hat{A}}^s\Mat{X}^s_{l}\,,
\end{equation}
where $\Mat{\hat{A}}^s=\Mat{D}^{-0.5}\Mat{A}^s\Mat{D}^{-0.5}$ denotes the normalized adjacency matrix without self loops and $l$ is the index of the propagation layer.
We let $\Mat{X}^s_{0}=\Mat{X}^s$, consisting of input user and item representations (e.g., $\Vector{e}_u$ and $\Vector{e}_i$).

After propagating for $L$ layers, we obtain multiple layer-wise representation matrices, i.e.,  $\Mat{\tilde{X}}^s=[\Mat{X}^s_0;\Mat{X}^s_1;\dots;\Mat{X}^s_L]$.
Since the representations of different layers capture different semantics, it is urgent to design a reasonable method to combine these representations. 
We implement five types of operations and validate user/item-specific GRUs~\cite{ChoMGBBSB14} can lead to good performance in general (see experimental results in Section~\ref{sec:ana}).

To be specific, we use $\Mat{\tilde{X}}^s_U$ to denote the user-specific representations in $\Mat{\tilde{X}}^s$ and it is analogous to $\Mat{\tilde{X}}^s_I$.
Consequently, we obtain the updated user representations $\Mat{\bar{X}}^s_U $ and item representations $\Mat{\bar{X}}^s_I$ by:
\begin{equation}\label{eq:GRU-FuseLayer}
    \Mat{\bar{X}}^s_U = \Normal{GRU}(\Mat{\tilde{X}}^s_U;\Theta^1_U)|_{L+1}\,,\quad\quad \Mat{\bar{X}}^s_I = \Normal{GRU}(\Mat{\tilde{X}}^s_I;\Theta^1_I)|_{L+1}\,,
\end{equation}
where $\Normal{GRU}(\cdot)|_{L+1}$ means returning the hidden representation from the recurrent step of $L+1$.
$\Theta^1_U$ and $\Theta^1_I$ are the learnable parameters for the respective GRU model.

Finally, to model the evolution of the temporal dynamics in the user and item sides, we further correlate the user and item representations across time slices by another two GRUs.
This is achieved by collecting time-sliced user representations and item representations to have $\Mat{\bar{X}}_U=[\Mat{\bar{X}}^1_U;\Mat{\bar{X}}^2_U;\dots;\Mat{\bar{X}}^T_U]$ and $\Mat{\bar{X}}_I=[\Mat{\bar{X}}^1_I;\Mat{\bar{X}}^2_I;\dots;\Mat{\bar{X}}^T_I]$.
Upon this, we perform recurrent computations to incorporate sequential information into the contextualized representations by:
\begin{equation}\label{eq:GRU-FuseTime}
    \Mat{H}_U = \Normal{GRU}(\Mat{\bar{X}}_U;\Theta^2_U)\,,\quad\quad \Mat{H}_I = \Normal{GRU}(\Mat{\bar{X}}_I;\Theta^2_I)\,,
\end{equation}
where $\Theta^2_U$ and $\Theta^2_I$ are the parameters of the above GRU models.

\subsection{Primary Sequential Recommendation Task}
Based on the proposed time-sliced graph neural networks, we have two sequences of representations for user $u$ and item $i$, i.e., $\Mat{H}_u=[\Vector{h}^1_u;\Vector{h}^2_u;\dots;\Vector{h}^{T}_u]$ and $\Mat{H}_i=[\Vector{h}^1_i;\Mat{h}^2_i;\dots;\Mat{h}^{T}_i]$.
As for predicting the potential interaction probability for the considered user-item pair, we combine their dynamic representations and static embeddings.
Specifically, we take the output of time-sliced graph neural networks, i.e., the last time-sliced user representations $\Vector{h}_u^T$ for user $u$ and item representations $\Vector{h}_i^T$ for item $i$, as the dynamic representations.
The representations mapped from IDs are regarded as static representations.
After concatenating them, multi-layer perceptrons (MLPs) are adopted to calculate the probability, which is given by:
\begin{equation}\label{eq:probability}
    \hat{y}_{ui} = \sigma\big(\Normal{MLPs}(\Vector{h}^{T}_u\oplus\Vector{h}^{T}_i\oplus\Vector{e}_u\oplus\Vector{e}_i;\Theta^{MLP})\big)\,,
\end{equation}
where we use ReLU as the middle-layered activation function.
$\sigma$ is the sigmoid function to let the value has a range $(0,1)$.
$\oplus$ means the concatenation operation.
$\Theta^{MLP}$ represents all the parameters involved in MLPs.

For training the model DRL-SRe, we formulate a sequential recommendation loss function w.r.t. the user-item pair $(u, i)$ based on the cross-entropy loss, which is given by:
\begin{equation}\label{eq:loss}
    \Loss_c = -\big(y_{ui}\log\hat{y}_{ui}+(1-y_{ui})\log(1-\hat{y}_{ui})\big)\,.
\end{equation}

\subsection{Auxiliary Temporal Prediction Task}
As discussed previously, although the timeline segmentation enables to model high-order and complex user-item relations through the time-sliced graph neural networks, the detailed temporal information is inevitably lost, which might cause the representations $\Mat{H}_u$ and $\Mat{H}_i$ to be sub-optimal.
To mitigate this issue, we introduce an auxiliary temporal prediction task based on temporal point process to compensate for the primary task.

Temporal point process takes the conditional intensity function $\lambda^*(t)$ as the most important component in capturing the continuous-time temporal dynamics. 
By convention, we use $^*$ in a function to indicate it is history-dependent. 
Within a short time interval $[t, t+dt)$, $\lambda^*(t)$ represents the occurrence rate of an event given the history $H_t$ and satisfies: $\lambda^*(t)dt = P\{c \in [t,t+dt]|H_t\}$. 
Based on this, the density function is given by:
\begin{equation}
    f^*(t) = \lambda^*(t)\exp(-\int_{t_j}^{t} \lambda^*(\epsilon)\, d\epsilon)
\end{equation}
where $t_j$ is the timestamp of the last event or a starting timestamp.

Under the situation of time-sliced user-item interaction graphs, we take the last interaction of a user or an item in a time slice as an event.
The corresponding time of the interaction is regarded as the event time.
Particularly, we use $[t_u^1,t_u^2,\cdots,t_u^T]$ and $[t_i^1,t_i^2,\cdots,t_i^T]$ to denote the detailed temporal information for user $u$ and item $i$ in different time slices, respectively.

Following the study~\cite{DuDTUGS16} that provides a well-designed conditional intensity function to derive an analytical form of the density function, we define the conditional intensity functions for user $u$ and item $i$ w.r.t. the $s$-th time slice as follows:
\begin{equation}
    \lambda^{*}_{u}(t) = \exp\big(\Vector{w_U} \Vector{h}^s_u+\omega_U(t-t^s_u)+b_U\big)\,,
\end{equation}
\begin{equation}
    \lambda^{*}_{i}(t) = \exp\big(\Vector{w_I} \Vector{h}^s_i+\omega_I(t-t^s_i)+b_I\big)\,,
\end{equation}
where $\Theta_U^3=\{\Vector{w_U},\omega_U,b_U\}$ and $\Theta_I^3=\{\Vector{w_I},\omega_I,b_I\}$ are the trainable parameters.
Through this manner, both the dynamic representations and temporal information are associated with intensities to characterize the continuous-time gaps over consecutive time slices from the dual sides.

The analytical form of the density functions could now be easily obtained.
For user $u$, it is defined as follows:
\begin{equation}\label{eq:density}
\begin{array}{l}
f^{*}_{u}(t) = \lambda^*_u(t)exp(-\int_{t^s_u}^{t} \lambda^*_u(\epsilon)\, d\epsilon) \\
     = \exp\{\Vector{w_U} \Vector{h}^s_u+\omega_U(t-t^s_u)+b_U+\frac{1}{\omega_U}\exp(\Vector{w_U}\Vector{h}^s_u + b_U) \\
     -\frac{1}{\omega_U}\exp(\Vector{w_U} \Vector{h}^s_u+\omega_U(t-t^s_u)+b_U)\}\,.
\end{array}
\end{equation}
Analogously, its density function could be derived for item $i$. 
Finally, the loss function of the auxiliary temporal prediction task is formulated as the negative joint log-likelihood of generating temporal sequences:
\begin{equation}
    \Loss_p = -\sum_{s=1}^{T-1}\log f^*_u(t^{s+1}_u|\Vector{h}^s_u)-\sum_{s=1}^{T-1}\log f^*_i(t^{s+1}_i|\Vector{h}^s_i)\,.
\end{equation}
Benefiting from this loss function, learning $\Vector{h}^s_u$ and $\Vector{h}^s_i$ is guided by the fine-grained continuous-time information.

\subsection{Model Training}
In the end, we unify the primary sequential recommendation task and the auxiliary temporal prediction task for jointly training DRL-SRe.
The hybrid objective function is defined as follows:
\begin{equation}\label{eq:final-loss}
    \Loss = \Loss_c+\beta\Loss_p\,,
\end{equation}
where $\beta$ is a hyper-parameter to control the relative effect of the auxiliary task.

The loss $\Loss$ is easy to be extended to a mini-batch setting where multiple user-item interactions are included.
L2 regularization and dropout strategies are commonly employed to alleviate the overfitting issue.
We use the Adam optimizer to learn the model parameters $\Theta=\{\Mat{E}_U, \Mat{E}_I, \Theta^{\{1,2,3\}}_U, \Theta^{\{1,2,3\}}_I, \Theta^{MLP}\}$.

Since the time-sliced graphs are only constructed for users and items having interactions in corresponding time slices, the number of the edges from all time-sliced graphs are of the same order of magnitude as the edge number of a global user-item interaction graph. 
Also, the complexity of sparse matrix multiplication depends on the number of edges in the Laplacian matrix.
As such, the computation does not cost much.

\section{Experiments}
This section first clarifies the experimental setups and then provides comprehensive experimental results, striving to answer the pivotal questions below:
\begin{itemize}[leftmargin=1.8em]
\item[\textbf{\texttt{Q1}.}] What are the results of the comparison between DRL-SRe and the state-of-the-art sequential recommendation models?

\item[\textbf{\texttt{Q2}.}] Does the performance of DRL-SRe suffer from a notable drop when removing any crucial component from the model?

\item[\textbf{\texttt{Q3}.}] How do alternative designs and hyper-parameter settings of DRL-SRe affect the final performance?
\end{itemize}

\begin{table}[!t]
\centering
\caption{Statistics of the datasets.}\label{tbl:stat}
\vspace{-1em}
\begin{tabular}{lccc}
\hline
\textbf{Dataset}  & Baby  & Yelp  & Netflix \\ \hline
 \# Users & 11,860 & 22,886 & 314,201 \\
 \# Items & 11,314 & 28,091 & 17,276 \\
 \# Interactions & 699,254 & 4,402,067 & 10,971,024 \\
 \# Time slices & 21 & 12 & 8 \\
 Time interval & 60 days & 30 days & 15 days \\
\hline
\end{tabular}
\end{table}

\begin{table*}[!t]
\centering
\caption{Main results w.r.t. HR@10, NDCG@10, and MRR for sequential recommendation on three datasets. The best and second-best performed methods in each metric are highlighted in “bold” and underline, respectively. Improv. denotes the relative improvement over the second-best results.}\label{tbl:performance-comp}
\vspace{-1.em}
\begin{tabular}{cccc|ccc|ccc} 
\hline
\multirow{2}*{Method}& \multicolumn{3}{c}{Baby}& \multicolumn{3}{c}{Yelp}&\multicolumn{3}{c}{Netflix}\\\cline{2-10}
&\multicolumn{1}{c}{HR@10} &\multicolumn{1}{c}{NDCG@10} &\multicolumn{1}{c}{MRR} &\multicolumn{1}{c}{HR@10} &\multicolumn{1}{c}{NDCG@10} &\multicolumn{1}{c}{MRR}
&\multicolumn{1}{c}{HR@10} &\multicolumn{1}{c}{NDCG@10} &\multicolumn{1}{c}{MRR} \\ \hline

GRU4Rec
&\multicolumn{1}{c}{0.7964} &\multicolumn{1}{c}{0.7530} &\multicolumn{1}{c}{0.7466} &\multicolumn{1}{c}{0.5677} &\multicolumn{1}{c}{0.4696} &\multicolumn{1}{c}{0.4505}
&\multicolumn{1}{c}{0.3514} &\multicolumn{1}{c}{0.2239} &\multicolumn{1}{c}{0.2047} \\

Caser
&\multicolumn{1}{c}{0.8024} &\multicolumn{1}{c}{0.7699} &\multicolumn{1}{c}{0.7665} &\multicolumn{1}{c}{0.5806} &\multicolumn{1}{c}{0.4753} &\multicolumn{1}{c}{0.4557}
&\multicolumn{1}{c}{0.2979} &\multicolumn{1}{c}{0.2050} &\multicolumn{1}{c}{0.1982} \\

SASRec
&\multicolumn{1}{c}{0.7996} &\multicolumn{1}{c}{0.7846} &\multicolumn{1}{c}{0.7866} &\multicolumn{1}{c}{0.6012} &\multicolumn{1}{c}{0.5148} &\multicolumn{1}{c}{0.5034}
&\multicolumn{1}{c}{0.3508} &\multicolumn{1}{c}{0.2450} &\multicolumn{1}{c}{0.2306} \\

TiSASRec
&\multicolumn{1}{c}{0.8084} &\multicolumn{1}{c}{\underline{0.7963}} &\multicolumn{1}{c}{\underline{ 0.8005 }}
&\multicolumn{1}{c}{0.6323} &\multicolumn{1}{c}{0.5407} &\multicolumn{1}{c}{0.5228}
&\multicolumn{1}{c}{0.3887} &\multicolumn{1}{c}{0.2642} &\multicolumn{1}{c}{0.2417} \\

ARNPP
&\multicolumn{1}{c}{0.8061} &\multicolumn{1}{c}{0.7829} &\multicolumn{1}{c}{0.7705}
&\multicolumn{1}{c}{0.6135} &\multicolumn{1}{c}{0.4904} &\multicolumn{1}{c}{0.4646}
&\multicolumn{1}{c}{0.3378} &\multicolumn{1}{c}{0.2319} &\multicolumn{1}{c}{0.2230} \\

LESSR
&\multicolumn{1}{c}{0.8104} &\multicolumn{1}{c}{0.7796} &\multicolumn{1}{c}{0.7772}
&\multicolumn{1}{c}{\underline{0.6581}} &\multicolumn{1}{c}{0.5784} &\multicolumn{1}{c}{0.5628}
&\multicolumn{1}{c}{0.3663} &\multicolumn{1}{c}{0.2560} &\multicolumn{1}{c}{0.2444} \\

RRN
&\multicolumn{1}{c}{\underline{ 0.8163 }} &\multicolumn{1}{c}{0.7604} &\multicolumn{1}{c}{0.7452} &\multicolumn{1}{c}{0.6258} &\multicolumn{1}{c}{0.5584} &\multicolumn{1}{c}{0.5490}
&\multicolumn{1}{c}{0.4084} &\multicolumn{1}{c}{0.2808} &\multicolumn{1}{c}{0.2698} \\

DEEMS
&\multicolumn{1}{c}{0.8141} &\multicolumn{1}{c}{0.7894} &\multicolumn{1}{c}{0.7879} &\multicolumn{1}{c}{0.6000} &\multicolumn{1}{c}{0.5593} &\multicolumn{1}{c}{0.5563}
&\multicolumn{1}{c}{0.4515} &\multicolumn{1}{c}{0.3117} &\multicolumn{1}{c}{0.3012} \\

SCoRe
&\multicolumn{1}{c}{0.8133} &\multicolumn{1}{c}{0.7951} &\multicolumn{1}{c}{0.7913} &\multicolumn{1}{c}{0.6494} &\multicolumn{1}{c}{\underline{0.5871}} &\multicolumn{1}{c}{\underline{0.5781}}
&\multicolumn{1}{c}{\underline{0.5339}} &\multicolumn{1}{c}{\underline{0.3659}} &\multicolumn{1}{c}{\underline{0.3295}} \\
\hline
\textbf{Ours}
&\multicolumn{1}{c}{\textbf{0.8423}} &\multicolumn{1}{c}{\textbf{0.8233}} &\multicolumn{1}{c}{\textbf{0.8237}} 
&\multicolumn{1}{c}{\textbf{0.6960}} &\multicolumn{1}{c}{\textbf{0.6471}} &\multicolumn{1}{c}{\textbf{0.6435}} 
&\multicolumn{1}{c}{\textbf{0.5742}} &\multicolumn{1}{c}{\textbf{0.4011}} &\multicolumn{1}{c}{\textbf{0.3607}}  \\
\hline
Improv.
&\multicolumn{1}{c}{3.19\%} &\multicolumn{1}{c}{3.39\%} &\multicolumn{1}{c}{2.90\%} &\multicolumn{1}{c}{5.76\%} &\multicolumn{1}{c}{10.22\%} &\multicolumn{1}{c}{11.31\%}
&\multicolumn{1}{c}{7.55\%} &\multicolumn{1}{c}{9.62\%} &\multicolumn{1}{c}{9.47\%} \\
\hline
\end{tabular}
\end{table*}

\subsection{Experimental Setups}

\subsubsection{Datasets} 
To evaluate the performance of all the models while ensuring reliability, we choose three datasets that are publicly available and with different origins, which are introduced as follows:

\textbf{Baby} is a category-specific subset extracted from the large public dataset~\cite{HeM16} of the e-commerce platform Amazon, ranging from May 1996 to July 2014.
This kind of dataset source is used in the relevant dual sequence model~\cite{WuGGWC19}.

\textbf{Yelp} is a review-based dataset that was released by Yelp in 2019.
It contains user ratings on businesses such as restaurants, bars, and spas, etc.
We regard the ratings with scores larger than 3 as positive feedback and select their corresponding user-item interactions to build the dataset.

\textbf{Netflix}\cite{bell2007lessons} is a commonly-used dataset from the Netflix contest which contains 100M ratings collected between November 1999 and December 2005.
This kind of dataset source is also used in sequential recommendation with dual sides~\cite{WuABSJ17}. 

To ensure the statistical significance of the datasets, we remove the users and items with less than 5 interactions.
The basic statistics of the three datasets are summarized in Table~\ref{tbl:stat}.
As shown in the line of ``Time slices'' in the table, we segment the whole timeline into $T$ time slices with an equal-length time interval --- shown in the line of ``Time interval'' --- for each dataset.
We use the user-item interactions coming from the first time slice to the $(T-2)$-th time slice as the training data.
The interactions in the $(T-1)$-th time slice are collected to constitute the validation data.
And the interactions in the last time slice are leveraged as the testing data.
This data setting is shared by all the adopted sequential recommendation models introduced later, so as to make fair comparisons.
It is worth noting that deep sequential recommendation models are usually formulated as predicting the next interacted item.
But they could be easily extended to train and predict the next several items~\cite{TangW18} that will be interacted with a target user in the next time slice.

\subsubsection{Evaluation Protocols}
We choose three widely used metrics in recommender systems: (1) HR@k (Hit Ratio@k) is the proportion of recommendation lists that have at least one positive item within top-k positions.
(2) NDCG@k (Normalized Discounted Cumulative Gain@k) is a position-aware ranking metric that assigns larger weights to the top positions.
As the positive items rank higher, the metric value becomes larger.
(3) MRR (Mean Reciprocal Rank) measures the relative position of the top-ranked positive item and takes a value of 1 if the positive item ranks at the first position.
As such, HR@k and MRR mainly focus on the first positive item, while NDCG@k considers a wider range of positive items.

Since it is too time-consuming to iterate over all user-item pairs to generate the complete sorting list and compute the above metrics, we follow the negative sampling strategy which is commonly observed in recommendation studies considering implicit feedback~\cite{HeLZNHC17,KangM18,QinRF-WSDM20}.
Specifically, for each ground-truth item, we randomly sample 100 negative items for metric computation. 
Without loss of generality, we show the results w.r.t. HR@10, NDCG@10, and MRR.

\subsubsection{Baselines}
We choose some representative sequential recommendation models that only consider the temporal dynamics in the user side as follows:
\begin{itemize}[leftmargin=*]
    \item[o] \textbf{GRU4Rec}~\cite{hidasi2015session}. This is a pioneering model that successfully applies recurrent neural networks to the sequential recommendation problem.
    It takes the interacted items for a target user as input to generate sequential recommendations. 
    
	\item[o] \textbf{Caser}~\cite{TangW18}. To see how convolutional neural networks (CNNs) behave on the used datasets, we choose Caser that combines CNNs and a latent factor model to learn users' sequential and general representations.
	
	\item[o] \textbf{SASRec}~\cite{KangM18}. SASRec is a well-performed model that heavily relies on self-attention mechanisms to identify important items from a user's behavior history.
	These important items affect user representations and finally determine the next-item prediction.
	
	\item[o] \textbf{TiSASRec}~\cite{LiWM20}. It is recently proposed by the same research group of SASRec and could be regarded as its enhancement.
	In particular, continuous time interval information between interacted items is encoded to facilitate the self-attention computation.

	\item[o] \textbf{ARNPP}~\cite{Liang20TNNLS}. ARNPP is a joint item and time prediction model that is empowered by temporal point process and attentive recurrent neural networks.
	Note that social relations utilized in the study~\cite{Liang20TNNLS} are not considered in this paper.
	
	\item[o] \textbf{LESSR}~\cite{ChenW20}. This model inherits the benefit of graph neural networks for session-based recommendation and additionally proposes an edge-order preserving aggregation strategy to capture the sequential order of interacted items.	
\end{itemize}

\begin{table*}[!t]
\centering
\caption{Ablation study of DRL-SRe.}\label{tbl:ablation}
\vspace{-1.em}
\begin{tabular}{l|ccc|ccc|ccc} 
\toprule
\multirow{2}*{\textbf{Method}}& \multicolumn{3}{c}{Baby}& \multicolumn{3}{c}{Yelp}&\multicolumn{3}{c}{Netflix}\\\cline{2-10}
&HR@10 &NDCG@10 &MRR &HR@10 &NDCG@10 &MRR &HR@10 &NDCG@10 &MRR \\\midrule[1.1pt]

\textbf{Ours} & \textbf{0.8423} & \textbf{0.8233}  & \textbf{0.8237}  & \textbf{0.6960}  & \textbf{0.6471} & \textbf{0.6435} & \textbf{0.5742} & \textbf{0.4011} & \textbf{0.3607}\\

w/o graph
& 0.8204 & 0.8118  & 0.8142 & 0.6774  & 0.6358 & 0.6343 & 0.5430 & 0.3766 & 0.3422 \\

w/o RNN (time slices)
& 0.8123 & 0.8055  & 0.8023 & 0.6439  & 0.6213 & 0.6132 & 0.5419 & 0.3674 & 0.3283 \\

w/o concat ID embedding
& 0.8343 & 0.8178  & 0.8172 & 0.6796  & 0.6355 & 0.6303 & 0.5582 & 0.3879 & 0.3544 \\

w/o auxiliary task
&0.8230 & 0.8157 & 0.8176 & 0.6646 & 0.6328 & 0.6249 & 0.5575 & 0.3729 & 0.3422 \\

- w/ time embedding
&0.8245 & 0.8102 & 0.8152 & 0.6623 & 0.6359 & 0.6230 & 0.5640 & 0.3912 & 0.3523 \\

\bottomrule
\end{tabular}
\end{table*}

We also take the following three well-known dual sequential recommendation models for comparisons:
\begin{itemize}[leftmargin=*]
	\item[o] \textbf{RRN}~\cite{WuABSJ17}. As the first model to address the dynamics in dual sides under the situation of sequential recommendation, it couples two RNNs to model the user temporal patterns and item temporal patterns, respectively.  
	
	\item[o] \textbf{DEEMS}~\cite{WuGGWC19}. DEEMS also relies on the architecture of the coupled RNNs and improves the objective function with a novel perspective of building dual prediction tasks.
	We use DEEMS-RNN for its slightly better performance.
	
	\item[o] \textbf{SCoRe}~\cite{QinRF-WSDM20}. This is the state-of-the-art dual sequence recommendation model that considers cross-neighbor relation modeling to better model relations between a target user and a candidate item.
	An interactive attention mechanism is adopted to gain an overall representation for the user or the item.
\end{itemize}

\subsubsection{Model Implementations}
We implement our model by Tensorflow and deploy it on a Linux server with GPUs of Nvidia GeForce GTX 2080 Ti (11G memory).
The model is learned in a mini-batch fashion with a size of 100.
For the adopted Adam optimizer, we set the learning rate to 5e-4 and keep the other hyper-parameters by default.
We add L2 regularization to the loss function (Equation~\ref{eq:loss}) by setting the regularization weight to 1e-4.
The embedding size of all the relevant models is fixed to 16 for ensuring fairness.
The number of layers used in MLP is set to 3.
The source code of this paper is available at \textbf{\url{https://github.com/weizhangltt/dual-recommend}}.

\subsection{Experimental Results}

\subsubsection{Performance Comparison (\textbf{\texttt{Q1}})}
Table~\ref{tbl:performance-comp} presents the overall performance of our model and all the adopted baselines, from which we have the following key observations:
\begin{itemize}[leftmargin=*]
\item Compared with other models, GRU4Rec achieves poor results on the three datasets.
This conforms to the expectation since only using the representation from the last time step of an RNN is insufficient.
The reason is that standard RNNs are not good at modeling long-range dependencies and have the forgetting issue.
As such, user preference information hidden in behavior sequences could not be effectively distilled. 

\item SASRec outperforms both the RNN-based model GRU4Rec and the CNN-based model Caser.
Such improvement might be attributed to the self-attention mechanism, which can assign larger weights to the important interacted items that will affect future user-item interactions. 

\item Compared to other sequential recommendation models that do not consider the dynamics of the item side, TiSASRec achieves better performance in most cases.
It makes sense as TiSASRec is an enhancement of SASRec by additionally considering the time interval information between the interacted items in behavior sequences.
Thanks to this, attention computation is further promoted. 

\item By comparing ARNPP with GRU4Rec, we could see the benefits brought by temporal point process.
By further investigating the performance of LESSR, we find it is even comparable with some dual sequence recommendation models in some cases.
This demonstrates the power of graph neural networks in modeling user-item interactions.

\item The three dual sequential recommendation models exhibit the relatively good performance from a whole perspective, demonstrating the positive effect of modeling the dual dynamics.
Among them, RRN is not as good as the other two models.
This might be caused by the limited ability of RNNs in modeling long-range dependency in a sequence.
DEEMS obtains better results than RRN, showing the positive contribution of introducing the dual prediction loss.
Thanks to the incorporation of local user-item interactions constrained to one sequence, SCoRe achieves the second-best results in many cases.

\item Our model DRL-SRe yields consistently better performance than all the baselines.
In particular, DRL-SRe improves the second-best performed models w.r.t. NDCG@10 by 3.39\%, 10.22\%, and 9.62\% on Baby, Yelp, and Netflix, respectively. 
This makes sense because: (1) DRL-SRe can distill knowledge from all the user-item interactions that happened in the past to benefit the dynamic user and item representation learning. 
(2) The introduction of the auxiliary temporal prediction task over consecutive time slices can significantly improve the sequential recommendation task, which is validated in Table~\ref{tbl:ablation}.

\end{itemize}

\begin{table*}[!t]
\centering
\caption{Results of model variants for DRL-SRe. Improvements over variants are statistically significant with p < 0.01.}\label{tbl:variants}
\vspace{-1.em}
\begin{tabular}{l|ccc|ccc|ccc} 
\toprule
\multirow{2}*{\textbf{Method}}& \multicolumn{3}{c}{Baby}& \multicolumn{3}{c}{Yelp}&\multicolumn{3}{c}{Netflix}\\\cline{2-10}
&HR@10 &NDCG@10 &MRR &HR@10 &NDCG@10 &MRR &HR@10 &NDCG@10 &MRR \\\midrule[1.1pt]

\textbf{Ours} & \textbf{0.8423} & \textbf{0.8233}  & \textbf{0.8237}  & \textbf{0.6960}  & \textbf{0.6471} & \textbf{0.6435} & \textbf{0.5742} & \textbf{0.4011} & \textbf{0.3607}\\

w/ global graph
& 0.7784 & 0.7373  & 0.7327 & 0.6042  & 0.5528 & 0.5399 & 0.5252 & 0.3473 & 0.3096 \\

w/ last graph
& 0.7625 & 0.6798  & 0.6624 & 0.6022  & 0.5107 & 0.4952 & 0.3676 & 0.2469 & 0.2281 \\

w/ user slices
& 0.8303 & 0.8010  & 0.7982 & 0.6624  & 0.6025 & 0.5933 & 0.3529 & 0.2286 & 0.2081 \\

w/ item slices
& 0.8282  & 0.8128 & 0.8123 & 0.6860  & 0.6345 & 0.6267 & 0.5311 & 0.3605 & 0.3086 \\
\hline
w/ concatenation
& 0.8321 & 0.8175  & 0.8177 & 0.6916  & 0.6440 & 0.6362  &0.5604 &0.3915 & 0.3541\\

w/ last-layer output
& 0.8347 & 0.8172  & 0.8167 & 0.6817  & 0.6306 & 0.6231  & 0.5666 & 0.3920 & 0.3374\\

w/ mean-pooling
& 0.8363 & 0.8181  & 0.8169 & 0.6774  & 0.6313 & 0.6253  & 0.5602 & 0.3871 & 0.3486\\

w/ single GRU
& 0.8325 & 0.8189  & 0.8204 & 0.6753  & 0.6348 & 0.6310 &0.5658 & 0.3837 & 0.3417 \\

\bottomrule
\end{tabular}
\end{table*}

\subsubsection{Ablation Study (\textbf{\texttt{Q2}})}
We further conduct an ablation study to validate the contributions of key components in DRL-SRe.
Specifically, (1) ``w/o graph'' represents removing graph neural networks and using simple mean-pooling to aggregate the neighbor information for users and items.
(2) ``w/o RNN (time slices)'' means not using Equation~\ref{eq:GRU-FuseTime} and performing a mean-pooling operation for the user/item representations obtained by time-sliced graph neural networks.
(3) ``w/o concat ID embedding'' simplifies Equation~\ref{eq:probability} by not concatenating the static user and item embeddings.
(4) ``w/o auxiliary task'' is equivalent to setting $\beta=0$ in Equation~\ref{eq:final-loss}, meaning not utilizing the auxiliary task. And ``- w/ time embedding'' leverages temporal embeddings for discretized time slots as model input instead of using the temporal prediction task.
 
Throughout the result analysis of the ablation study shown in Table~\ref{tbl:ablation},
we observe that:
\begin{itemize}[leftmargin=*]
\item ``w/o graph'' suffers severe performance degradation.
It validates the crucial role of time-sliced user-item interaction graphs for characterizing the dynamics of users and items.
This is intuitive since learning from full user-item interactions in a time slice could lead to accurate time-sliced user and item representations.

\item ``w/o RNN (time)'' also sees its significant performance drop.
This phenomenon reveals that explicitly correlating time-sliced user representations and item representations is indispensable. 
The reason might be that RNNs could introduce sequential information that each time-sliced representation does not have. 

\item The concatenation of static embeddings has a positive impact by seeing the results of ``w/o concat ID embedding''. 

\item ``w/o auxiliary task'' is obviously inferior to the full model DRL-SRe, verifying that using temporal point process to capture fine-grained temporal information is advantageous for deriving effective dynamic representations.
Moreover, simply feeding time embeddings into model input is not as good as leveraging the temporal prediction task for sequential recommendation.

\end{itemize}

\subsubsection{Analysis of Model Alternatives (\textbf{\texttt{Q3}})}\label{sec:ana}
To deeply investigate why the proposed model works, we design several alternatives of the model:
(1) ``w/ global graph'' replaces the multiple time-sliced graphs with a global user-item interaction graph.
(2) ``w/ last graph'' replaces the multiple time-sliced graphs with a last user-item interaction graph.
(3) ``w/ user slices'' only uses the sequences w.r.t. interacted users to do final prediction, i.e., $\Vector{h}^{T}_u\oplus\Vector{e}_u\oplus\Vector{e}_i$ in Equation~\ref{eq:probability}.
(4) ``w/ item slices'' only adopts the sequences w.r.t. interacted items to perform prediction, i.e., $\Vector{h}^{T}_i\oplus\Vector{e}_u\oplus\Vector{e}_i$ in Equation~\ref{eq:probability}.

Besides, we investigate some different operations to fuse layer-wise representations from time-sliced graph neural networks, as shown in Equation~\ref{eq:GRU-FuseLayer}.
``w/ concatenation'' means concatenating the layer-wise representations.
``w/ last-layer output'' denotes only using the representation of the last layer.
``w/ mean-pooling'' uses mean-pooling to summarize the layer-wise representations.
And ``w/ single GRU'' does not differentiate the user side and the item side by only employing a single GRU.

From the results shown in Table~\ref{tbl:variants}, we find that:
\begin{itemize}[leftmargin=*]
\item In the first part of the table, the performance degradation of ``w/ global graph'' is significant. 
It is intuitive since only using a global user-item interaction graph is unable to capture the temporal dynamics of users and items, although the ability to encode enriched user-item interactions by GNN is maintained.
Similarly, ``w/ last graph'' is degraded heavily, which reflects the great necessity of modeling long behavior sequences for the studied task.
Moreover, by evaluating the results of ``w/ user slices'' and ``w/ item slices'', we could conclude that learning from dual dynamics is profitable. 

\item In the second part of the table, the four alternatives perform worse than using user/item-specific GRUs in DRL-SRe.
In particular, the comparison between ``w/ single GRU'' and the full model validates that differentiating the dual sides of users and items consistently boosts the recommendation performance.

\end{itemize}

\subsubsection{More Discussions (\textbf{\texttt{Q3}})} 
This part provides some more discussions about how some hyper-parameters affect DRL-SRe, including hyper-parameter $\beta$ and the propagation layer number of GNNs.

\begin{figure}[!h]
    \centering
    \subfloat
    {
    \includegraphics[width=0.49\linewidth]{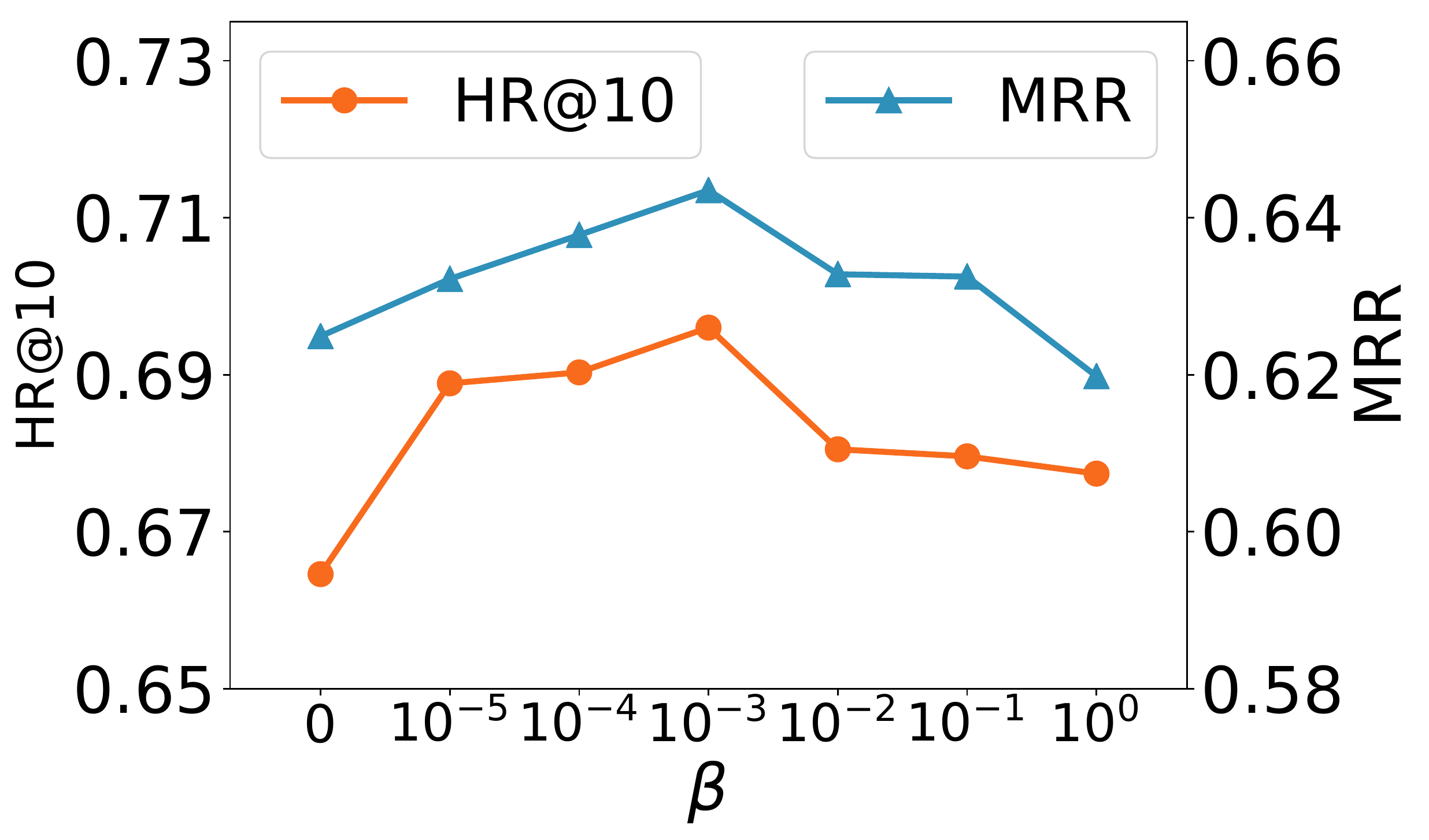}\label{fig:yelp_pp}
    \includegraphics[width=0.49\linewidth]{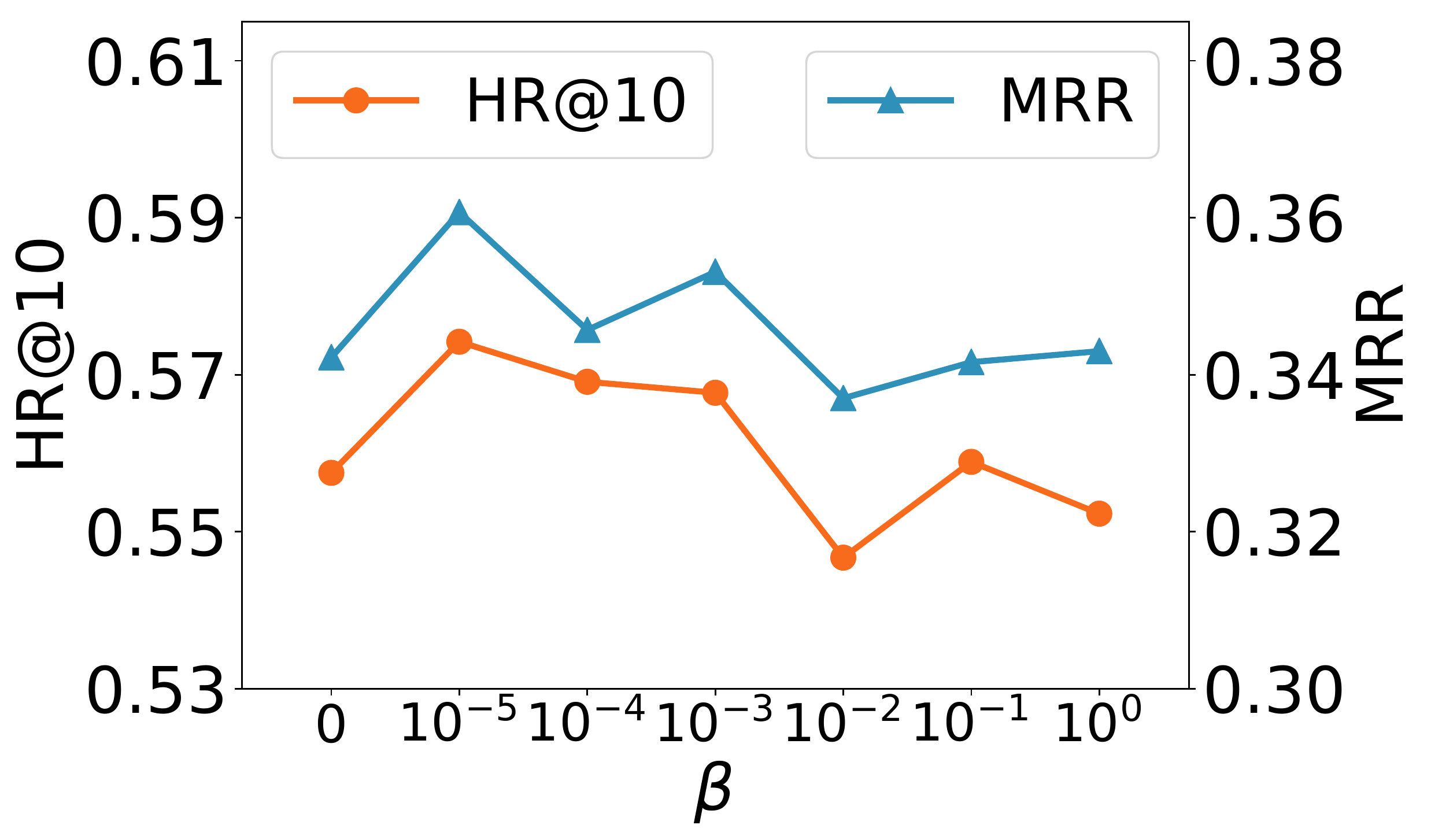}\label{fig:netflix_pp}

    }
    \caption{Result variation with different $\beta$ on Yelp (left) and Netflix (right).}
    \label{fig:beta_var}
\end{figure}

\noindent \textbf{Effect of hyper-parameter $\beta$.}
To analyze the influence of hyper-parameter $\beta$, we vary $\beta$ in range of $\{0,10^{-5},10^{-4},10^{-3},10^{-2},10^{-1},10^{0}\}$ and illustrate the recommendation performance changing curve of DRL-SRe on the Yelp and Netflix datasets in figure~\ref{fig:beta_var}.
We can observe that better results are achieved when setting $\beta$ in a suitable middle value range and too large values might even hurt the performance.

\begin{figure}[!h]
    \centering
    \subfloat
    {\includegraphics[width=0.33\linewidth]{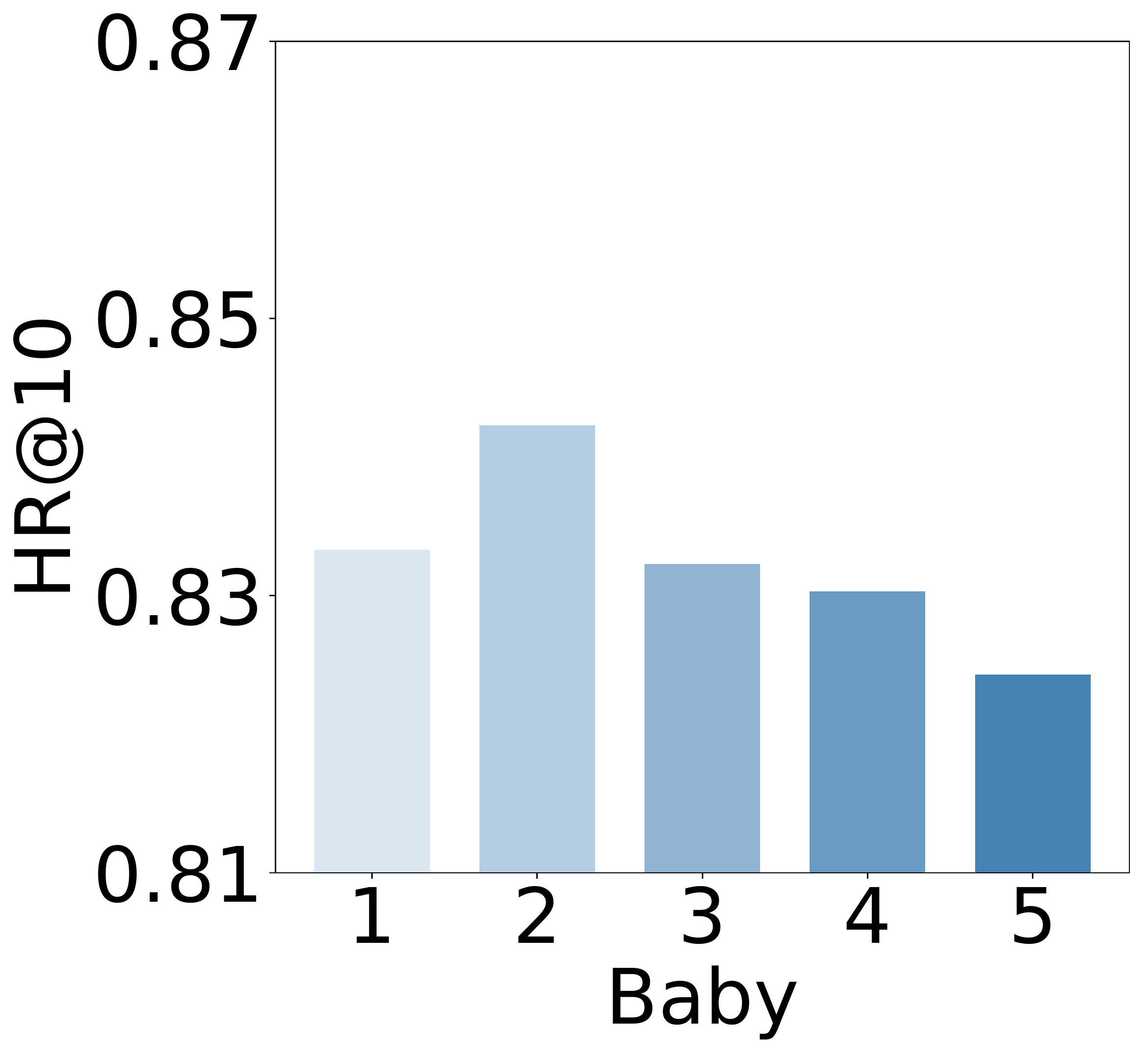}\label{fig:baby_layer}
    \includegraphics[width=0.33\linewidth]{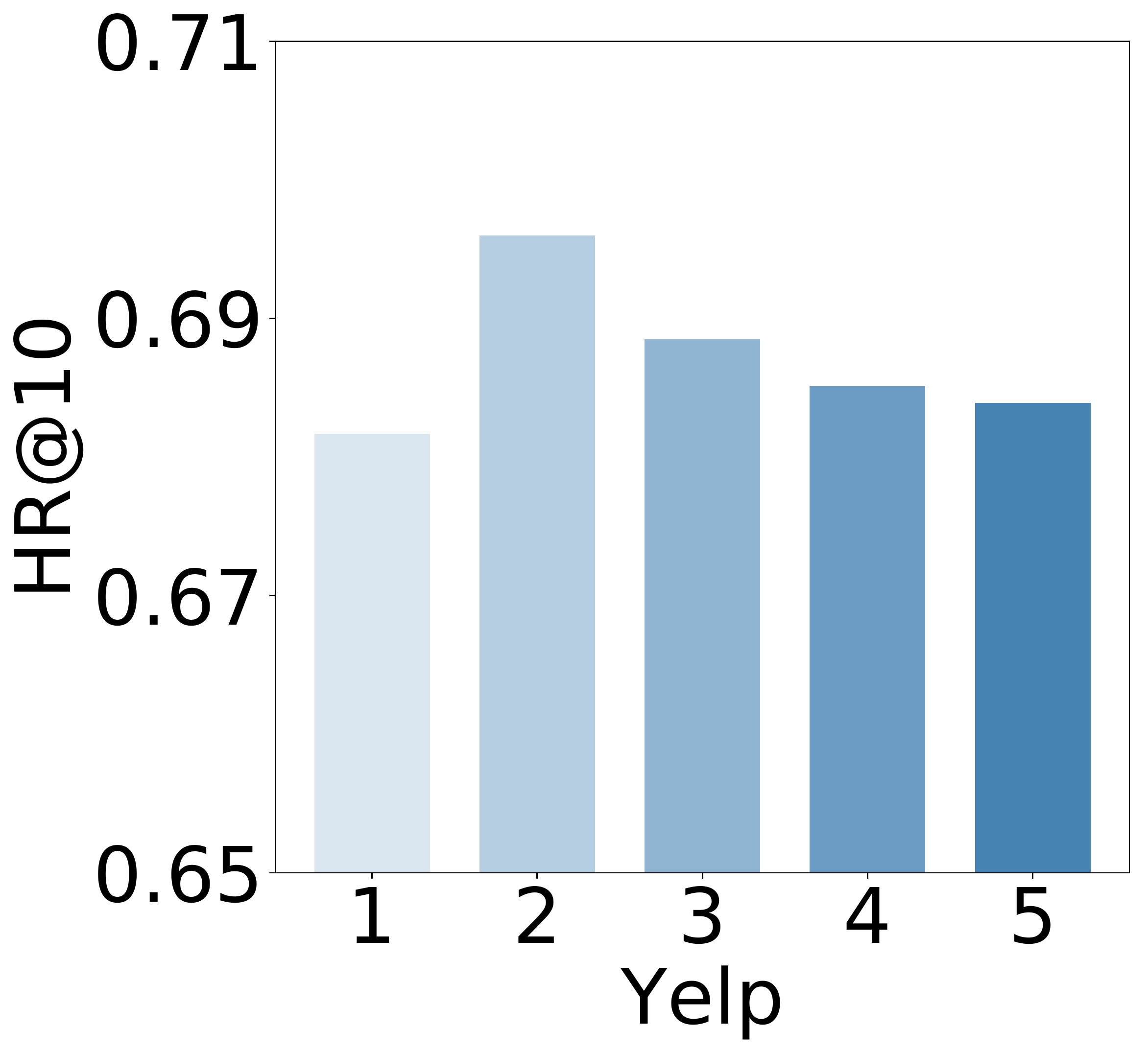}\label{fig:yelp_layer}
    \includegraphics[width=0.33\linewidth]{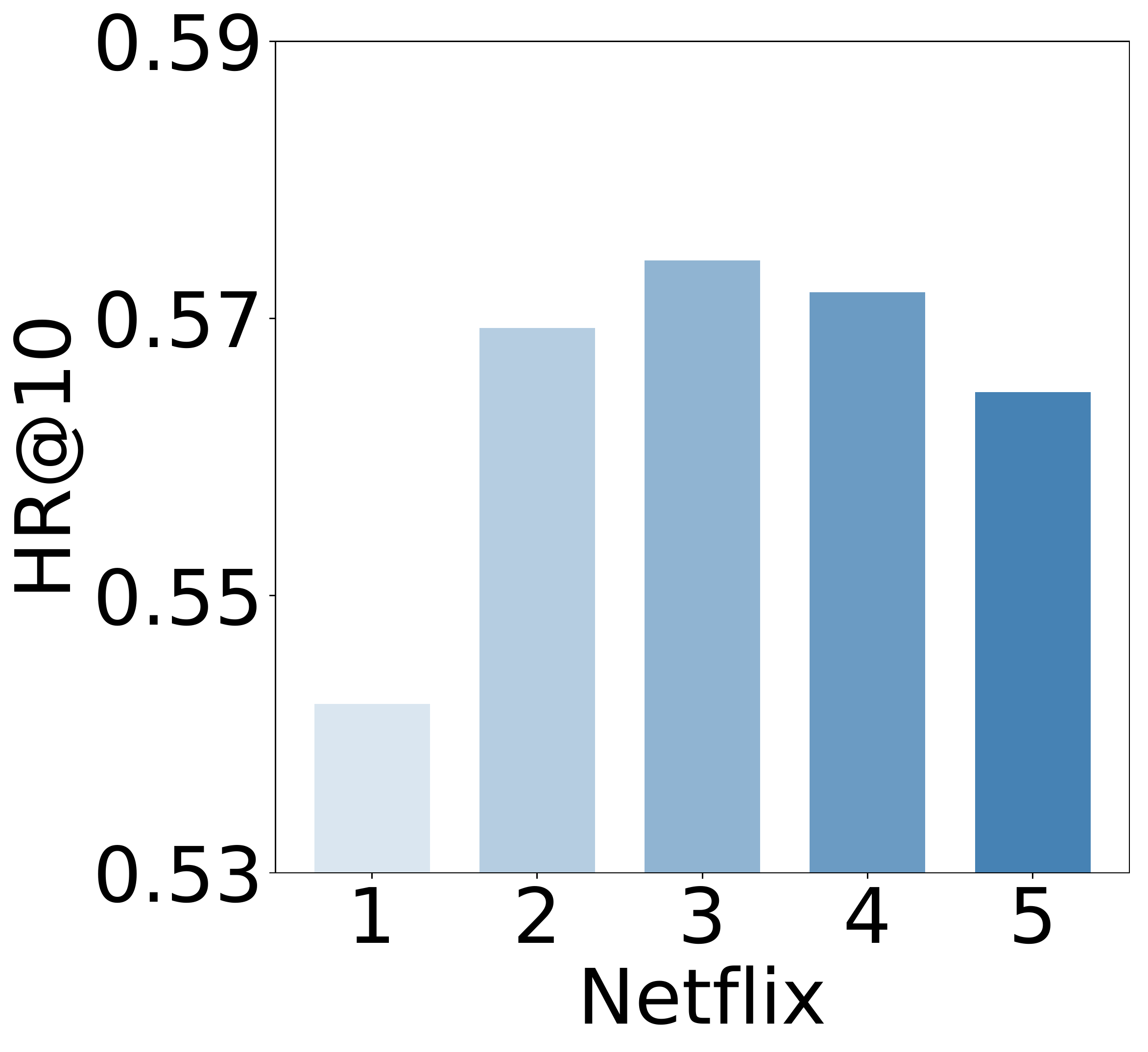}\label{fig:netflix_layer}
    }
    \caption{Result variation when increasing the number of propagation layers (shown in x-axis).}
    \label{fig:layer_num}
\end{figure}

\noindent \textbf{Effect of layer numbers in GNN.}
To investigate whether our model can benefit from the multi-layer propagation, we vary the number of GNN layers in the range of $\{1,2,3,4,5\}$.
Figure~\ref{fig:layer_num} depicts the performance trend by HR@10.
When increasing the layer number from 1 to 2, DRL-SRe achieves consistently better performance on the three datasets, indicating that introducing neighbor information through representation propagation is profitable.
However, the continual increase of the layer number even harms the performance.
This might be caused by the over-smoothing and overfitting issues.

\section{Conclusion}\label{sec:conclu}
This paper studies the sequential recommendation problem by modeling the dual temporal dynamics for both the user and item sides.
The novel model DRL-SRe is proposed, which innovatively develops the time-sliced graph neural networks to learn dynamic user and item representations, and introduces the auxiliary temporal prediction task to compensate for the primary sequential recommendation task based on temporal point process.
The comprehensive experiments conducted on three public and large datasets validate the superiority of DRL-SRe and the effectiveness of its main components.

\begin{acks}
This work was supported in part by National Natural Science Foundation of China under Grant (No. 62072182) and the Fundamental Research Funds for the Central Universities.
\end{acks}

\bibliographystyle{ACM-Reference-Format}
\bibliography{references}


\begin{thebibliography}{46}


\ifx \showCODEN    \undefined \def \showCODEN     #1{\unskip}     \fi
\ifx \showDOI      \undefined \def \showDOI       #1{#1}\fi
\ifx \showISBNx    \undefined \def \showISBNx     #1{\unskip}     \fi
\ifx \showISBNxiii \undefined \def \showISBNxiii  #1{\unskip}     \fi
\ifx \showISSN     \undefined \def \showISSN      #1{\unskip}     \fi
\ifx \showLCCN     \undefined \def \showLCCN      #1{\unskip}     \fi
\ifx \shownote     \undefined \def \shownote      #1{#1}          \fi
\ifx \showarticletitle \undefined \def \showarticletitle #1{#1}   \fi
\ifx \showURL      \undefined \def \showURL       {\relax}        \fi
\providecommand\bibfield[2]{#2}
\providecommand\bibinfo[2]{#2}
\providecommand\natexlab[1]{#1}
\providecommand\showeprint[2][]{arXiv:#2}

\bibitem[\protect\citeauthoryear{Bell and Koren}{Bell and Koren}{2007}]%
        {bell2007lessons}
\bibfield{author}{\bibinfo{person}{Robert~M Bell} {and} \bibinfo{person}{Yehuda
  Koren}.} \bibinfo{year}{2007}\natexlab{}.
\newblock \showarticletitle{Lessons from the netflix prize challenge}.
\newblock \bibinfo{journal}{\emph{SIGKDD Explorations Newsletter}}
  (\bibinfo{year}{2007}), \bibinfo{pages}{75--79}.
\newblock


\bibitem[\protect\citeauthoryear{Chang, Liu, Wen, Li, Fang, Song, and Qi}{Chang
  et~al\mbox{.}}{2020}]%
        {ChangLW0FS020}
\bibfield{author}{\bibinfo{person}{Xiaofu Chang}, \bibinfo{person}{Xuqin Liu},
  \bibinfo{person}{Jianfeng Wen}, \bibinfo{person}{Shuang Li},
  \bibinfo{person}{Yanming Fang}, \bibinfo{person}{Le Song}, {and}
  \bibinfo{person}{Yuan Qi}.} \bibinfo{year}{2020}\natexlab{}.
\newblock \showarticletitle{Continuous-Time Dynamic Graph Learning via Neural
  Interaction Processes}. In \bibinfo{booktitle}{\emph{CIKM}}.
  \bibinfo{pages}{145--154}.
\newblock


\bibitem[\protect\citeauthoryear{Chen and Wong}{Chen and Wong}{2020}]%
        {ChenW20}
\bibfield{author}{\bibinfo{person}{Tianwen Chen} {and}
  \bibinfo{person}{Raymond~Chi{-}Wing Wong}.} \bibinfo{year}{2020}\natexlab{}.
\newblock \showarticletitle{Handling Information Loss of Graph Neural Networks
  for Session-based Recommendation}. In \bibinfo{booktitle}{\emph{SIGKDD}}.
  \bibinfo{pages}{1172--1180}.
\newblock


\bibitem[\protect\citeauthoryear{Chen, Long, Cong, and Li}{Chen
  et~al\mbox{.}}{2020}]%
        {ChenLCL20}
\bibfield{author}{\bibinfo{person}{Yile Chen}, \bibinfo{person}{Cheng Long},
  \bibinfo{person}{Gao Cong}, {and} \bibinfo{person}{Chenliang Li}.}
  \bibinfo{year}{2020}\natexlab{}.
\newblock \showarticletitle{Context-aware Deep Model for Joint Mobility and
  Time Prediction}. In \bibinfo{booktitle}{\emph{WSDM}}.
  \bibinfo{pages}{106--114}.
\newblock


\bibitem[\protect\citeauthoryear{Cho, van Merrienboer, G{\"{u}}l{\c{c}}ehre,
  Bahdanau, Bougares, Schwenk, and Bengio}{Cho et~al\mbox{.}}{2014}]%
        {ChoMGBBSB14}
\bibfield{author}{\bibinfo{person}{Kyunghyun Cho}, \bibinfo{person}{Bart van
  Merrienboer}, \bibinfo{person}{{\c{C}}aglar G{\"{u}}l{\c{c}}ehre},
  \bibinfo{person}{Dzmitry Bahdanau}, \bibinfo{person}{Fethi Bougares},
  \bibinfo{person}{Holger Schwenk}, {and} \bibinfo{person}{Yoshua Bengio}.}
  \bibinfo{year}{2014}\natexlab{}.
\newblock \showarticletitle{Learning Phrase Representations using {RNN}
  Encoder-Decoder for Statistical Machine Translation}. In
  \bibinfo{booktitle}{\emph{EMNLP}}. \bibinfo{pages}{1724--1734}.
\newblock


\bibitem[\protect\citeauthoryear{Du, Dai, Trivedi, Upadhyay, Gomez{-}Rodriguez,
  and Song}{Du et~al\mbox{.}}{2016}]%
        {DuDTUGS16}
\bibfield{author}{\bibinfo{person}{Nan Du}, \bibinfo{person}{Hanjun Dai},
  \bibinfo{person}{Rakshit Trivedi}, \bibinfo{person}{Utkarsh Upadhyay},
  \bibinfo{person}{Manuel Gomez{-}Rodriguez}, {and} \bibinfo{person}{Le Song}.}
  \bibinfo{year}{2016}\natexlab{}.
\newblock \showarticletitle{Recurrent Marked Temporal Point Processes:
  Embedding Event History to Vector}. In \bibinfo{booktitle}{\emph{SIGKDD}}.
  \bibinfo{pages}{1555--1564}.
\newblock


\bibitem[\protect\citeauthoryear{Durbin and Watson}{Durbin and Watson}{1971}]%
        {Durbin1971Spectra}
\bibfield{author}{\bibinfo{person}{J. Durbin} {and} \bibinfo{person}{G.~S.
  Watson}.} \bibinfo{year}{1971}\natexlab{}.
\newblock \showarticletitle{Spectra of some self-exciting and mutually exciting
  point processes}.
\newblock \bibinfo{journal}{\emph{Biometrika}} \bibinfo{volume}{58},
  \bibinfo{number}{1} (\bibinfo{year}{1971}), \bibinfo{pages}{83--90}.
\newblock


\bibitem[\protect\citeauthoryear{Fan, Ma, Li, He, Zhao, Tang, and Yin}{Fan
  et~al\mbox{.}}{2019}]%
        {Fan0LHZTY19}
\bibfield{author}{\bibinfo{person}{Wenqi Fan}, \bibinfo{person}{Yao Ma},
  \bibinfo{person}{Qing Li}, \bibinfo{person}{Yuan He},
  \bibinfo{person}{Yihong~Eric Zhao}, \bibinfo{person}{Jiliang Tang}, {and}
  \bibinfo{person}{Dawei Yin}.} \bibinfo{year}{2019}\natexlab{}.
\newblock \showarticletitle{Graph Neural Networks for Social Recommendation}.
  In \bibinfo{booktitle}{\emph{WWW}}. \bibinfo{pages}{417--426}.
\newblock


\bibitem[\protect\citeauthoryear{Feng, Li, Zhang, Sun, Meng, Guo, and Jin}{Feng
  et~al\mbox{.}}{2018}]%
        {FengLZSMGJ18}
\bibfield{author}{\bibinfo{person}{Jie Feng}, \bibinfo{person}{Yong Li},
  \bibinfo{person}{Chao Zhang}, \bibinfo{person}{Funing Sun},
  \bibinfo{person}{Fanchao Meng}, \bibinfo{person}{Ang Guo}, {and}
  \bibinfo{person}{Depeng Jin}.} \bibinfo{year}{2018}\natexlab{}.
\newblock \showarticletitle{DeepMove: Predicting Human Mobility with
  Attentional Recurrent Networks}. In \bibinfo{booktitle}{\emph{WWW}}.
  \bibinfo{pages}{1459--1468}.
\newblock


\bibitem[\protect\citeauthoryear{He and McAuley}{He and McAuley}{2016}]%
        {HeM16}
\bibfield{author}{\bibinfo{person}{Ruining He} {and} \bibinfo{person}{Julian~J.
  McAuley}.} \bibinfo{year}{2016}\natexlab{}.
\newblock \showarticletitle{Ups and Downs: Modeling the Visual Evolution of
  Fashion Trends with One-Class Collaborative Filtering}. In
  \bibinfo{booktitle}{\emph{WWW}}. \bibinfo{pages}{507--517}.
\newblock


\bibitem[\protect\citeauthoryear{He, Deng, Wang, Li, Zhang, and Wang}{He
  et~al\mbox{.}}{2020}]%
        {HeGCN2020}
\bibfield{author}{\bibinfo{person}{Xiangnan He}, \bibinfo{person}{Kuan Deng},
  \bibinfo{person}{Xiang Wang}, \bibinfo{person}{Yan Li},
  \bibinfo{person}{Yong{-}Dong Zhang}, {and} \bibinfo{person}{Meng Wang}.}
  \bibinfo{year}{2020}\natexlab{}.
\newblock \showarticletitle{LightGCN: Simplifying and Powering Graph
  Convolution Network for Recommendation}. In
  \bibinfo{booktitle}{\emph{SIGIR}}. \bibinfo{pages}{639--648}.
\newblock


\bibitem[\protect\citeauthoryear{He, Liao, Zhang, Nie, Hu, and Chua}{He
  et~al\mbox{.}}{2017}]%
        {HeLZNHC17}
\bibfield{author}{\bibinfo{person}{Xiangnan He}, \bibinfo{person}{Lizi Liao},
  \bibinfo{person}{Hanwang Zhang}, \bibinfo{person}{Liqiang Nie},
  \bibinfo{person}{Xia Hu}, {and} \bibinfo{person}{Tat{-}Seng Chua}.}
  \bibinfo{year}{2017}\natexlab{}.
\newblock \showarticletitle{Neural Collaborative Filtering}. In
  \bibinfo{booktitle}{\emph{WWW}}. \bibinfo{pages}{173--182}.
\newblock


\bibitem[\protect\citeauthoryear{Hidasi, Karatzoglou, Baltrunas, and
  Tikk}{Hidasi et~al\mbox{.}}{2015}]%
        {hidasi2015session}
\bibfield{author}{\bibinfo{person}{Bal{\'a}zs Hidasi},
  \bibinfo{person}{Alexandros Karatzoglou}, \bibinfo{person}{Linas Baltrunas},
  {and} \bibinfo{person}{Domonkos Tikk}.} \bibinfo{year}{2015}\natexlab{}.
\newblock \showarticletitle{Session-based recommendations with recurrent neural
  networks}.
\newblock \bibinfo{journal}{\emph{ICLR}} (\bibinfo{year}{2015}).
\newblock


\bibitem[\protect\citeauthoryear{Kang and McAuley}{Kang and McAuley}{2018}]%
        {KangM18}
\bibfield{author}{\bibinfo{person}{Wang{-}Cheng Kang} {and}
  \bibinfo{person}{Julian~J. McAuley}.} \bibinfo{year}{2018}\natexlab{}.
\newblock \showarticletitle{Self-Attentive Sequential Recommendation}. In
  \bibinfo{booktitle}{\emph{ICDM}}. \bibinfo{pages}{197--206}.
\newblock


\bibitem[\protect\citeauthoryear{Kouki, Fountalis, Vasiloglou, Cui, Liberty,
  and Jadda}{Kouki et~al\mbox{.}}{2020}]%
        {KoukiFVCLJ20}
\bibfield{author}{\bibinfo{person}{Pigi Kouki}, \bibinfo{person}{Ilias
  Fountalis}, \bibinfo{person}{Nikolaos Vasiloglou}, \bibinfo{person}{Xiquan
  Cui}, \bibinfo{person}{Edo Liberty}, {and} \bibinfo{person}{Khalifeh~Al
  Jadda}.} \bibinfo{year}{2020}\natexlab{}.
\newblock \showarticletitle{From the lab to production: {A} case study of
  session-based recommendations in the home-improvement domain}. In
  \bibinfo{booktitle}{\emph{RecSys}}. \bibinfo{pages}{140--149}.
\newblock


\bibitem[\protect\citeauthoryear{Kumar, Zhang, and Leskovec}{Kumar
  et~al\mbox{.}}{2019}]%
        {KumarZL19}
\bibfield{author}{\bibinfo{person}{Srijan Kumar}, \bibinfo{person}{Xikun
  Zhang}, {and} \bibinfo{person}{Jure Leskovec}.}
  \bibinfo{year}{2019}\natexlab{}.
\newblock \showarticletitle{Predicting Dynamic Embedding Trajectory in Temporal
  Interaction Networks}. In \bibinfo{booktitle}{\emph{SIGKDD}}.
  \bibinfo{pages}{1269--1278}.
\newblock


\bibitem[\protect\citeauthoryear{Li, Wang, and McAuley}{Li
  et~al\mbox{.}}{2020a}]%
        {LiWM20}
\bibfield{author}{\bibinfo{person}{Jiacheng Li}, \bibinfo{person}{Yujie Wang},
  {and} \bibinfo{person}{Julian~J. McAuley}.} \bibinfo{year}{2020}\natexlab{a}.
\newblock \showarticletitle{Time Interval Aware Self-Attention for Sequential
  Recommendation}. In \bibinfo{booktitle}{\emph{WSDM}}.
  \bibinfo{pages}{322--330}.
\newblock


\bibitem[\protect\citeauthoryear{Li, Zhang, Wu, Liu, Wang, and Yu}{Li
  et~al\mbox{.}}{2020b}]%
        {LiZWLWY20}
\bibfield{author}{\bibinfo{person}{Xiaohan Li}, \bibinfo{person}{Mengqi Zhang},
  \bibinfo{person}{Shu Wu}, \bibinfo{person}{Zheng Liu}, \bibinfo{person}{Liang
  Wang}, {and} \bibinfo{person}{Philip~S. Yu}.}
  \bibinfo{year}{2020}\natexlab{b}.
\newblock \showarticletitle{Dynamic Graph Collaborative Filtering}. In
  \bibinfo{booktitle}{\emph{ICDM}}. \bibinfo{pages}{322--331}.
\newblock


\bibitem[\protect\citeauthoryear{Li, Zhao, Liu, Huang, Mei, and Chen}{Li
  et~al\mbox{.}}{2018}]%
        {LiZLHMC18}
\bibfield{author}{\bibinfo{person}{Zhi Li}, \bibinfo{person}{Hongke Zhao},
  \bibinfo{person}{Qi Liu}, \bibinfo{person}{Zhenya Huang},
  \bibinfo{person}{Tao Mei}, {and} \bibinfo{person}{Enhong Chen}.}
  \bibinfo{year}{2018}\natexlab{}.
\newblock \showarticletitle{Learning from History and Present: Next-item
  Recommendation via Discriminatively Exploiting User Behaviors}. In
  \bibinfo{booktitle}{\emph{SIGKDD}}. \bibinfo{pages}{1734--1743}.
\newblock


\bibitem[\protect\citeauthoryear{Lian, Wu, Ge, Xie, and Chen}{Lian
  et~al\mbox{.}}{2020}]%
        {LianWG0C20}
\bibfield{author}{\bibinfo{person}{Defu Lian}, \bibinfo{person}{Yongji Wu},
  \bibinfo{person}{Yong Ge}, \bibinfo{person}{Xing Xie}, {and}
  \bibinfo{person}{Enhong Chen}.} \bibinfo{year}{2020}\natexlab{}.
\newblock \showarticletitle{Geography-Aware Sequential Location
  Recommendation}. In \bibinfo{booktitle}{\emph{SIGKDD}}.
  \bibinfo{pages}{2009--2019}.
\newblock


\bibitem[\protect\citeauthoryear{Liang and Zhang}{Liang and Zhang}{2020}]%
        {Liang20TNNLS}
\bibfield{author}{\bibinfo{person}{Wenwei Liang} {and} \bibinfo{person}{Wei
  Zhang}.} \bibinfo{year}{2020}\natexlab{}.
\newblock \showarticletitle{Learning Social Relations and Spatiotemporal
  Trajectories for Next Check-in Inference}.
\newblock \bibinfo{journal}{\emph{IEEE TNNLS}} (\bibinfo{year}{2020}).
\newblock


\bibitem[\protect\citeauthoryear{Luo, Zhang, Yang, Bo, Yang, Li, Qie, and
  Ye}{Luo et~al\mbox{.}}{2020}]%
        {LuoZYBYLQY20}
\bibfield{author}{\bibinfo{person}{Wenjuan Luo}, \bibinfo{person}{Han Zhang},
  \bibinfo{person}{Xiaodi Yang}, \bibinfo{person}{Lin Bo},
  \bibinfo{person}{Xiaoqing Yang}, \bibinfo{person}{Zang Li},
  \bibinfo{person}{Xiaohu Qie}, {and} \bibinfo{person}{Jieping Ye}.}
  \bibinfo{year}{2020}\natexlab{}.
\newblock \showarticletitle{Dynamic Heterogeneous Graph Neural Network for
  Real-time Event Prediction}. In \bibinfo{booktitle}{\emph{SIGKDD}}.
  \bibinfo{pages}{3213--3223}.
\newblock


\bibitem[\protect\citeauthoryear{Pi, Bian, Zhou, Zhu, and Gai}{Pi
  et~al\mbox{.}}{2019}]%
        {PiBZZG19}
\bibfield{author}{\bibinfo{person}{Qi Pi}, \bibinfo{person}{Weijie Bian},
  \bibinfo{person}{Guorui Zhou}, \bibinfo{person}{Xiaoqiang Zhu}, {and}
  \bibinfo{person}{Kun Gai}.} \bibinfo{year}{2019}\natexlab{}.
\newblock \showarticletitle{Practice on Long Sequential User Behavior Modeling
  for Click-Through Rate Prediction}. In \bibinfo{booktitle}{\emph{SIGKDD}},
  \bibfield{editor}{\bibinfo{person}{Ankur Teredesai}, \bibinfo{person}{Vipin
  Kumar}, \bibinfo{person}{Ying Li}, \bibinfo{person}{R{\'{o}}mer Rosales},
  \bibinfo{person}{Evimaria Terzi}, {and} \bibinfo{person}{George Karypis}}
  (Eds.). \bibinfo{pages}{2671--2679}.
\newblock


\bibitem[\protect\citeauthoryear{Qin, Ren, Fang, Zhang, and Yu}{Qin
  et~al\mbox{.}}{2020}]%
        {QinRF-WSDM20}
\bibfield{author}{\bibinfo{person}{Jiarui Qin}, \bibinfo{person}{Kan Ren},
  \bibinfo{person}{Yuchen Fang}, \bibinfo{person}{Weinan Zhang}, {and}
  \bibinfo{person}{Yong Yu}.} \bibinfo{year}{2020}\natexlab{}.
\newblock \showarticletitle{Sequential Recommendation with Dual Side
  Neighbor-based Collaborative Relation Modeling}. In
  \bibinfo{booktitle}{\emph{WSDM}}, \bibfield{editor}{\bibinfo{person}{James
  Caverlee}, \bibinfo{person}{Xia~(Ben) Hu}, \bibinfo{person}{Mounia Lalmas},
  {and} \bibinfo{person}{Wei Wang}} (Eds.). \bibinfo{pages}{465--473}.
\newblock


\bibitem[\protect\citeauthoryear{Qiu, Huang, Li, and Yin}{Qiu
  et~al\mbox{.}}{2020}]%
        {QiuHLY20}
\bibfield{author}{\bibinfo{person}{Ruihong Qiu}, \bibinfo{person}{Zi Huang},
  \bibinfo{person}{Jingjing Li}, {and} \bibinfo{person}{Hongzhi Yin}.}
  \bibinfo{year}{2020}\natexlab{}.
\newblock \showarticletitle{Exploiting Cross-session Information for
  Session-based Recommendation with Graph Neural Networks}.
\newblock \bibinfo{journal}{\emph{ACM TOIS}} \bibinfo{volume}{38},
  \bibinfo{number}{3} (\bibinfo{year}{2020}), \bibinfo{pages}{22:1--22:23}.
\newblock


\bibitem[\protect\citeauthoryear{Quadrana, Cremonesi, and Jannach}{Quadrana
  et~al\mbox{.}}{2018}]%
        {QuadranaCJ18}
\bibfield{author}{\bibinfo{person}{Massimo Quadrana}, \bibinfo{person}{Paolo
  Cremonesi}, {and} \bibinfo{person}{Dietmar Jannach}.}
  \bibinfo{year}{2018}\natexlab{}.
\newblock \showarticletitle{Sequence-Aware Recommender Systems}.
\newblock \bibinfo{journal}{\emph{{ACM} Comput. Surv.}} \bibinfo{volume}{51},
  \bibinfo{number}{4} (\bibinfo{year}{2018}), \bibinfo{pages}{66:1--66:36}.
\newblock


\bibitem[\protect\citeauthoryear{Quadrana, Karatzoglou, Hidasi, and
  Cremonesi}{Quadrana et~al\mbox{.}}{2017}]%
        {quadrana2017personalizing}
\bibfield{author}{\bibinfo{person}{Massimo Quadrana},
  \bibinfo{person}{Alexandros Karatzoglou}, \bibinfo{person}{Bal{\'a}zs
  Hidasi}, {and} \bibinfo{person}{Paolo Cremonesi}.}
  \bibinfo{year}{2017}\natexlab{}.
\newblock \showarticletitle{Personalizing session-based recommendations with
  hierarchical recurrent neural networks}. In
  \bibinfo{booktitle}{\emph{RecSys}}. \bibinfo{pages}{130--137}.
\newblock


\bibitem[\protect\citeauthoryear{Sankar, Wu, Gou, Zhang, and Yang}{Sankar
  et~al\mbox{.}}{2020}]%
        {SankarWGZY20}
\bibfield{author}{\bibinfo{person}{Aravind Sankar}, \bibinfo{person}{Yanhong
  Wu}, \bibinfo{person}{Liang Gou}, \bibinfo{person}{Wei Zhang}, {and}
  \bibinfo{person}{Hao Yang}.} \bibinfo{year}{2020}\natexlab{}.
\newblock \showarticletitle{DySAT: Deep Neural Representation Learning on
  Dynamic Graphs via Self-Attention Networks}. In
  \bibinfo{booktitle}{\emph{WSDM}}. \bibinfo{pages}{519--527}.
\newblock


\bibitem[\protect\citeauthoryear{Shi, Larson, and Hanjalic}{Shi
  et~al\mbox{.}}{2014}]%
        {ShiLH14}
\bibfield{author}{\bibinfo{person}{Yue Shi}, \bibinfo{person}{Martha~A.
  Larson}, {and} \bibinfo{person}{Alan Hanjalic}.}
  \bibinfo{year}{2014}\natexlab{}.
\newblock \showarticletitle{Collaborative Filtering beyond the User-Item
  Matrix: {A} Survey of the State of the Art and Future Challenges}.
\newblock \bibinfo{journal}{\emph{{ACM} Comput. Surv.}} \bibinfo{volume}{47},
  \bibinfo{number}{1} (\bibinfo{year}{2014}), \bibinfo{pages}{3:1--3:45}.
\newblock


\bibitem[\protect\citeauthoryear{Tang and Wang}{Tang and Wang}{2018}]%
        {TangW18}
\bibfield{author}{\bibinfo{person}{Jiaxi Tang} {and} \bibinfo{person}{Ke
  Wang}.} \bibinfo{year}{2018}\natexlab{}.
\newblock \showarticletitle{Personalized Top-N Sequential Recommendation via
  Convolutional Sequence Embedding}. In \bibinfo{booktitle}{\emph{WSDM}},
  \bibfield{editor}{\bibinfo{person}{Yi~Chang}, \bibinfo{person}{Chengxiang
  Zhai}, \bibinfo{person}{Yan Liu}, {and} \bibinfo{person}{Yoelle Maarek}}
  (Eds.). \bibinfo{pages}{565--573}.
\newblock


\bibitem[\protect\citeauthoryear{Tang, Liu, Shah, Shi, Mitra, and Wang}{Tang
  et~al\mbox{.}}{2020}]%
        {TangLSSMW20}
\bibfield{author}{\bibinfo{person}{Xianfeng Tang}, \bibinfo{person}{Yozen Liu},
  \bibinfo{person}{Neil Shah}, \bibinfo{person}{Xiaolin Shi},
  \bibinfo{person}{Prasenjit Mitra}, {and} \bibinfo{person}{Suhang Wang}.}
  \bibinfo{year}{2020}\natexlab{}.
\newblock \showarticletitle{Knowing your {FATE:} Friendship, Action and
  Temporal Explanations for User Engagement Prediction on Social Apps}. In
  \bibinfo{booktitle}{\emph{SIGKDD}}. \bibinfo{pages}{2269--2279}.
\newblock


\bibitem[\protect\citeauthoryear{Vass{\o}y, Ruocco, de~Souza~da Silva, and
  Aune}{Vass{\o}y et~al\mbox{.}}{2019}]%
        {VassoyRSA19}
\bibfield{author}{\bibinfo{person}{Bj{\o}rnar Vass{\o}y},
  \bibinfo{person}{Massimiliano Ruocco}, \bibinfo{person}{Eliezer de~Souza~da
  Silva}, {and} \bibinfo{person}{Erlend Aune}.}
  \bibinfo{year}{2019}\natexlab{}.
\newblock \showarticletitle{Time is of the Essence: {A} Joint Hierarchical
  {RNN} and Point Process Model for Time and Item Predictions}. In
  \bibinfo{booktitle}{\emph{WSDM}}. \bibinfo{pages}{591--599}.
\newblock


\bibitem[\protect\citeauthoryear{Vaswani, Shazeer, Parmar, Uszkoreit, Jones,
  Gomez, Kaiser, and Polosukhin}{Vaswani et~al\mbox{.}}{2017}]%
        {VaswaniSPUJGKP17}
\bibfield{author}{\bibinfo{person}{Ashish Vaswani}, \bibinfo{person}{Noam
  Shazeer}, \bibinfo{person}{Niki Parmar}, \bibinfo{person}{Jakob Uszkoreit},
  \bibinfo{person}{Llion Jones}, \bibinfo{person}{Aidan~N. Gomez},
  \bibinfo{person}{Lukasz Kaiser}, {and} \bibinfo{person}{Illia Polosukhin}.}
  \bibinfo{year}{2017}\natexlab{}.
\newblock \showarticletitle{Attention is All you Need}. In
  \bibinfo{booktitle}{\emph{NIPS}}. \bibinfo{pages}{5998--6008}.
\newblock


\bibitem[\protect\citeauthoryear{Wang, Zhang, Wang, Zhao, Li, Xie, and
  Guo}{Wang et~al\mbox{.}}{2019b}]%
        {WangZWZLXG19}
\bibfield{author}{\bibinfo{person}{Hongwei Wang}, \bibinfo{person}{Fuzheng
  Zhang}, \bibinfo{person}{Jialin Wang}, \bibinfo{person}{Miao Zhao},
  \bibinfo{person}{Wenjie Li}, \bibinfo{person}{Xing Xie}, {and}
  \bibinfo{person}{Minyi Guo}.} \bibinfo{year}{2019}\natexlab{b}.
\newblock \showarticletitle{Exploring High-Order User Preference on the
  Knowledge Graph for Recommender Systems}.
\newblock \bibinfo{journal}{\emph{TOIS}} \bibinfo{volume}{37},
  \bibinfo{number}{3} (\bibinfo{year}{2019}), \bibinfo{pages}{32:1--32:26}.
\newblock


\bibitem[\protect\citeauthoryear{Wang, Zhang, Liu, Liu, Zhang, Lin, and
  Zha}{Wang et~al\mbox{.}}{2020b}]%
        {WangZLLZLZ20}
\bibfield{author}{\bibinfo{person}{Wen Wang}, \bibinfo{person}{Wei Zhang},
  \bibinfo{person}{Shukai Liu}, \bibinfo{person}{Qi Liu}, \bibinfo{person}{Bo
  Zhang}, \bibinfo{person}{Leyu Lin}, {and} \bibinfo{person}{Hongyuan Zha}.}
  \bibinfo{year}{2020}\natexlab{b}.
\newblock \showarticletitle{Beyond Clicks: Modeling Multi-Relational Item Graph
  for Session-Based Target Behavior Prediction}. In
  \bibinfo{booktitle}{\emph{WWW}}. \bibinfo{pages}{3056--3062}.
\newblock


\bibitem[\protect\citeauthoryear{Wang, Zhang, Rao, Qiu, Zhang, Lin, and
  Zha}{Wang et~al\mbox{.}}{2020c}]%
        {Wang0RQ0LZ20}
\bibfield{author}{\bibinfo{person}{Wen Wang}, \bibinfo{person}{Wei Zhang},
  \bibinfo{person}{Jun Rao}, \bibinfo{person}{Zhijie Qiu}, \bibinfo{person}{Bo
  Zhang}, \bibinfo{person}{Leyu Lin}, {and} \bibinfo{person}{Hongyuan Zha}.}
  \bibinfo{year}{2020}\natexlab{c}.
\newblock \showarticletitle{Group-Aware Long- and Short-Term Graph
  Representation Learning for Sequential Group Recommendation}. In
  \bibinfo{booktitle}{\emph{SIGIR}}. \bibinfo{pages}{1449--1458}.
\newblock


\bibitem[\protect\citeauthoryear{Wang, He, Wang, Feng, and Chua}{Wang
  et~al\mbox{.}}{2019a}]%
        {Wang0WFC19}
\bibfield{author}{\bibinfo{person}{Xiang Wang}, \bibinfo{person}{Xiangnan He},
  \bibinfo{person}{Meng Wang}, \bibinfo{person}{Fuli Feng}, {and}
  \bibinfo{person}{Tat{-}Seng Chua}.} \bibinfo{year}{2019}\natexlab{a}.
\newblock \showarticletitle{Neural Graph Collaborative Filtering}. In
  \bibinfo{booktitle}{\emph{SIGIR}}. \bibinfo{pages}{165--174}.
\newblock


\bibitem[\protect\citeauthoryear{Wang, Wei, Cong, Li, Mao, and Qiu}{Wang
  et~al\mbox{.}}{2020a}]%
        {Wang0CLMQ20}
\bibfield{author}{\bibinfo{person}{Ziyang Wang}, \bibinfo{person}{Wei Wei},
  \bibinfo{person}{Gao Cong}, \bibinfo{person}{Xiao{-}Li Li},
  \bibinfo{person}{Xianling Mao}, {and} \bibinfo{person}{Minghui Qiu}.}
  \bibinfo{year}{2020}\natexlab{a}.
\newblock \showarticletitle{Global Context Enhanced Graph Neural Networks for
  Session-based Recommendation}. In \bibinfo{booktitle}{\emph{SIGIR}}.
  \bibinfo{pages}{169--178}.
\newblock


\bibitem[\protect\citeauthoryear{Wu, Ahmed, Beutel, Smola, and Jing}{Wu
  et~al\mbox{.}}{2017}]%
        {WuABSJ17}
\bibfield{author}{\bibinfo{person}{Chao{-}Yuan Wu}, \bibinfo{person}{Amr
  Ahmed}, \bibinfo{person}{Alex Beutel}, \bibinfo{person}{Alexander~J. Smola},
  {and} \bibinfo{person}{How Jing}.} \bibinfo{year}{2017}\natexlab{}.
\newblock \showarticletitle{Recurrent Recommender Networks}. In
  \bibinfo{booktitle}{\emph{WSDM}}. \bibinfo{pages}{495--503}.
\newblock


\bibitem[\protect\citeauthoryear{Wu, Cao, Cheung, and Hamilton}{Wu
  et~al\mbox{.}}{2020}]%
        {WuCCH20}
\bibfield{author}{\bibinfo{person}{Jiapeng Wu}, \bibinfo{person}{Meng Cao},
  \bibinfo{person}{Jackie Chi~Kit Cheung}, {and} \bibinfo{person}{William~L.
  Hamilton}.} \bibinfo{year}{2020}\natexlab{}.
\newblock \showarticletitle{TeMP: Temporal Message Passing for Temporal
  Knowledge Graph Completion}. In \bibinfo{booktitle}{\emph{EMNLP}}.
  \bibinfo{pages}{5730--5746}.
\newblock


\bibitem[\protect\citeauthoryear{Wu, Gao, Gao, Weng, and Chen}{Wu
  et~al\mbox{.}}{2019a}]%
        {WuGGWC19}
\bibfield{author}{\bibinfo{person}{Qitian Wu}, \bibinfo{person}{Yirui Gao},
  \bibinfo{person}{Xiaofeng Gao}, \bibinfo{person}{Paul Weng}, {and}
  \bibinfo{person}{Guihai Chen}.} \bibinfo{year}{2019}\natexlab{a}.
\newblock \showarticletitle{Dual Sequential Prediction Models Linking
  Sequential Recommendation and Information Dissemination}. In
  \bibinfo{booktitle}{\emph{SIGKDD}}. \bibinfo{pages}{447--457}.
\newblock


\bibitem[\protect\citeauthoryear{Wu, Tang, Zhu, Wang, Xie, and Tan}{Wu
  et~al\mbox{.}}{2019b}]%
        {sr_gnn}
\bibfield{author}{\bibinfo{person}{Shu Wu}, \bibinfo{person}{Yuyuan Tang},
  \bibinfo{person}{Yanqiao Zhu}, \bibinfo{person}{Liang Wang},
  \bibinfo{person}{Xing Xie}, {and} \bibinfo{person}{Tieniu Tan}.}
  \bibinfo{year}{2019}\natexlab{b}.
\newblock \showarticletitle{Session-based recommendation with graph neural
  networks}. In \bibinfo{booktitle}{\emph{AAAI}}. \bibinfo{pages}{346--353}.
\newblock


\bibitem[\protect\citeauthoryear{Ying, He, Chen, Eksombatchai, Hamilton, and
  Leskovec}{Ying et~al\mbox{.}}{2018}]%
        {YingHCEHL18}
\bibfield{author}{\bibinfo{person}{Rex Ying}, \bibinfo{person}{Ruining He},
  \bibinfo{person}{Kaifeng Chen}, \bibinfo{person}{Pong Eksombatchai},
  \bibinfo{person}{William~L. Hamilton}, {and} \bibinfo{person}{Jure
  Leskovec}.} \bibinfo{year}{2018}\natexlab{}.
\newblock \showarticletitle{Graph Convolutional Neural Networks for Web-Scale
  Recommender Systems}. In \bibinfo{booktitle}{\emph{SIGKDD}},
  \bibfield{editor}{\bibinfo{person}{Yike Guo} {and} \bibinfo{person}{Faisal
  Farooq}} (Eds.). \bibinfo{pages}{974--983}.
\newblock


\bibitem[\protect\citeauthoryear{Zhang, Yao, Sun, and Tay}{Zhang
  et~al\mbox{.}}{2019}]%
        {ZhangYST19}
\bibfield{author}{\bibinfo{person}{Shuai Zhang}, \bibinfo{person}{Lina Yao},
  \bibinfo{person}{Aixin Sun}, {and} \bibinfo{person}{Yi Tay}.}
  \bibinfo{year}{2019}\natexlab{}.
\newblock \showarticletitle{Deep Learning Based Recommender System: {A} Survey
  and New Perspectives}.
\newblock \bibinfo{journal}{\emph{{ACM} Comput. Surv.}} \bibinfo{volume}{52},
  \bibinfo{number}{1} (\bibinfo{year}{2019}), \bibinfo{pages}{5:1--5:38}.
\newblock


\bibitem[\protect\citeauthoryear{Zhang, Dai, Xu, Feng, Wang, Bian, Wang, and
  Liu}{Zhang et~al\mbox{.}}{2014}]%
        {ZhangDXFWBWL14}
\bibfield{author}{\bibinfo{person}{Yuyu Zhang}, \bibinfo{person}{Hanjun Dai},
  \bibinfo{person}{Chang Xu}, \bibinfo{person}{Jun Feng},
  \bibinfo{person}{Taifeng Wang}, \bibinfo{person}{Jiang Bian},
  \bibinfo{person}{Bin Wang}, {and} \bibinfo{person}{Tie{-}Yan Liu}.}
  \bibinfo{year}{2014}\natexlab{}.
\newblock \showarticletitle{Sequential Click Prediction for Sponsored Search
  with Recurrent Neural Networks}. In \bibinfo{booktitle}{\emph{AAAI}}.
  \bibinfo{pages}{1369--1375}.
\newblock


\bibitem[\protect\citeauthoryear{Zhou, Cui, Zhang, Yang, Liu, and Sun}{Zhou
  et~al\mbox{.}}{2018}]%
        {ZhouGNN18}
\bibfield{author}{\bibinfo{person}{Jie Zhou}, \bibinfo{person}{Ganqu Cui},
  \bibinfo{person}{Zhengyan Zhang}, \bibinfo{person}{Cheng Yang},
  \bibinfo{person}{Zhiyuan Liu}, {and} \bibinfo{person}{Maosong Sun}.}
  \bibinfo{year}{2018}\natexlab{}.
\newblock \showarticletitle{Graph Neural Networks: {A} Review of Methods and
  Applications}.
\newblock \bibinfo{journal}{\emph{CoRR}}  \bibinfo{volume}{abs/1812.08434}
  (\bibinfo{year}{2018}).
\newblock


\end{thebibliography}

\end{document}